\documentclass[iop, revtex4]{emulateapj}

\usepackage{natbib}
\usepackage{graphicx}
\usepackage{latexsym}
\usepackage{amsmath}
\usepackage{float}

\slugcomment{Draft v8.0}

\shorttitle{NIR Variability in Tr 37}
\shortauthors{Meng et al. 2018}

\begin{document}

\title{Near-Infrared Variability of Low Mass Stars in IC 1396A and Tr 37}

\author{Huan Y. A. Meng\altaffilmark{1}, G. H. Rieke\altaffilmark{1}, Jinyoung Serena Kim\altaffilmark{1}, Aurora Sicilia-Aguilar\altaffilmark{2}, N. J. G. Cross\altaffilmark{3}, Taran Esplin\altaffilmark{1}, L. M. Rebull\altaffilmark{4}, Klaus W. Hodapp\altaffilmark{5} }
\altaffiltext{1}{Steward Observatory, Department of Astronomy, University of Arizona, 933 North Cherry Avenue, Tucson, AZ 85721}
\altaffiltext{2}{SUPA, School of Science and Engineering, University of Dundee, Nethergate, Dundee DD1 4HN, UK}
\altaffiltext{3}{Wide-Field Astronomy Unit, Institute for Astronomy, School of Physics and Astronomy, University of Edinburgh,
Royal Observatory, Blackford Hill, Edinburgh EH9 3HJ, UK}
\altaffiltext{4}{Spitzer Science Center, Infrared Processing and Analysis Center, California Institute of Technology, 1200 E. California Blvd., Pasadena, CA 91125}
\altaffiltext{5} {Institute for Astronomy, University of Hawaii, 640 N. Aohoku Place, Hilo, HI 96720, USA}

\email{hyameng@lpl.arizona.edu}

\begin{abstract}
We have monitored nearly a square degree in IC 1396A/Tr 37 over 21 epochs extending over 2014 $-$ 2016 for sources variable in the $JHK$ bands. In our data, $65 \pm 8$ \% of previously identified cluster members show variations, compared with $\lesssim0.3$\% of field stars. We identify 119 members of Tr 37 on the basis of variability, forming an unbiased sample down to the brown dwarf regime. The $K$-band luminosity function in Tr 37 is similar to that of IC 348 but shifted to somewhat brighter values, implying that the K- and M-type members of Tr 37 are younger than those in IC 348. We introduce methods to classify the causes of variability, based on behavior in the color-color and color-magnitude diagrams. Accretion hot spots cause larger variations at $J$ than at $K$ with substantial scatter in the diagrams; there are at least a dozen, with the most active resembling EXors.  Eleven sources are probably dominated by intervention of dust clumps in their circumstellar disks with color behavior indicating the presence of grains larger than for interstellar dust, presumably due to grain growth in their disks.  Thirteen sources have larger variations at $K$ than at $J$ or $H$. For 11 of them, the temperature fitted to the variable component is very close to 2000K, suggesting that the changes in output are caused by turbulence at the inner rim of the circumstellar disk exposing previously protected populations of grains. 
\end{abstract}

\keywords{protoplanetary disks --- infrared: planetary systems --- infrared: stars --- stars: pre-main sequence --- open clusters and associations: individual (Tr 37)}

\section{Introduction} \label{sec:intro}
%
Variability is a very common characteristic of young stellar objects (YSOs) and young sub-stellar objects
(YSSOs).  However, its prevalence only became apparent with systematic surveys
conducted with, for example, the 2-Micron All Sky Survey (2MASS) \citep{skrutskie2006} and with {\it Spitzer} \citep{werner2004}. The underlying causes for this behavior
are only partially understood, but can include variable accretion rates, magnetic activity,
flares, hot or cold starspots, and the effects of circumstellar disks such as changes in extinction or disk emission. 
 A better understanding of the variation
patterns can therefore give insight to many aspects of YSO behavior, including fundamental ones such as accretion 
and the structure of protoplanetary disks.

Optical and infrared (IR) variability studies have also been used as a powerful tool to
identify candidate YSOs.  At faint limits the colors of reddened background stars can mimic
faint cluster members. This issue can be combated with multiple observations to
identify members by variability, since only a tiny fraction of field stars are variable \citep[e.g.,][] {morales-calderon2009, pietrukowicz2009, wolk2013a, wolk2013b}.

To use variability as a tool to probe YSOs requires systematic surveys extending over long time
baselines.  For example, the YSOVAR program in the warm {\it Spitzer} mission (PI J. Stauffer)
monitored selected regions intensively for about a month twice a year (due to visibility constraints) \citep{rebull2014}.
In the Orion nebula cluster (ONC) this effort found that 77\% of disked stars and 44\% of
the weak-lined T Tauri Stars (WTTS) showed variations \citep{morales-calderon2011}. \citet{rice2015} used the Wide Field Camera (WFCAM) on 
the United Kingdom Infrared Telescope (UKIRT) to carry out 
a very extensive $JHK$ study of the same region, finding 1203 variable stars. 
\citet{rice2012, wolk2013a, wolk2013b} also used WFCAM to monitor Cyg OB7 over 1.5 years showing that 83\% of known YSOs
are variable and less than 2\% of the field stars are. They found about 60 short period (a few days)
variables, a similar number of stochastic variables, a number of stars with long periods (20 - 60 days),
and about 25 with unclassified variations. Other examples of near-infrared variability surveys include Orion A by \citet{carpenter2001}
using 2MASS data and $\rho$ Oph by \citet{deoliveria2008} with UKIRT/WFCAM.

The number of such studies is modest, but they have proven to give unique insights in areas like young star gas accretion and the structure of protoplanetary disks.  
To complement them, we have carried out
a multi-epoch deep near-IR imaging variability survey of the star forming region IC 1396A/Tr 37 using
UKIRT/WFCAM. Our target region covers a range
of ages from $<$ 0.5 Myr to 10 Myr, and we have monitored this region with 21 epochs of observation 
extending over nearly two years. 
We are able to obtain accurate variability and color information
down to $K\sim17 $ mag,  allowing a deep search for very low mass 
cluster members down into the brown dwarf regime.

%
IC~1396A is a bright-rimmed globule complex, also known as ``the Elephant Trunk Nebula",
illuminated by HD~206267, a Trapezium-like O star system (O6.5V + O9V), which is a part of
Cepheus OB2 association, and is at a distance of  945$^{+90}_{-73}$ pc from Gaia DR2 measurements \citep{gaia2018, sicilia-aguilar2019}.  
This value is consistent within the errors with the result from main sequence fitting of the intermediate mass 
($\sim$ A-type) stars, $870 \pm 80$ pc \citep{contreras2002}, 
particularly given a possible small bias in the Gaia results toward large distances \citep{xu2019}.
IC~1396A is at the western edge of the Tr~37 cluster \citep{marschall1987, platais1998}.
\citet{sicilia-aguilar2004, sicilia-aguilar2005,  sicilia-aguilar2006a, sicilia-aguilar2013} have conducted observations of
Tr~37 using optical photometric and spectroscopic data as well as {\it Spitzer} IRAC and MIPS in the IR 
to identify and study the young stellar population and circumstellar disk evolution. YSOs in Tr 37 have also been studied in the X-ray using the {\it Chandra} X-ray
Observatory \citep{getman2012} and  far-IR using the {\it Herschel} Space Observatory \citep{sicilia-aguilar2015}. From these studies, the Cep~OB2 association harbors multiple generations of star forming regions \citep{sicilia-aguilar2005, sicilia-aguilar2006b, sicilia-aguilar2019, morales-calderon2009}: NGC~7160 (10-12~Myr), Tr~37 (3-4~Myr), IC 1396A and
IC 1396N ($\lesssim$1~Myr). It provides an ideal laboratory to study YSO disk evolution under an external UV radiation
environment in multiple evolutionary phases, since the typical lifetimes of gaseous disks 
around low mass stars are $\lesssim$3 $-$ 5~Myr \citep[e.g.,][]{ribas2015, meng2017}

\citet{sicilia-aguilar2005} presented a list of optically variable sources in this region based on about a week of monitoring, 
and \citet{barentsen2011} presented $r$ and $i$ variable sources from ``several'' nights of their 
IPHAS survey, which covered a 7~deg$^2$ region in Cep~OB2 that included IC 1396.
Mid-IR variability in the IC~1396A region has been studied with {\it Spitzer} by \citet{morales-calderon2009}. 
This study conducted two modes of observations:
a 14 day time span with twice a day cadence as well as 7 hours of continuous ``staring" mode
observations, and found that more than 50\% of the YSOs were variable. \citet{scholz2010} 
monitored IC 1369W with WFCAM in JHK intensively for three nights, discovering two eclipsing binaries and 
eight periodic variables.

\begin{figure*}
\epsscale{1}
\plotone{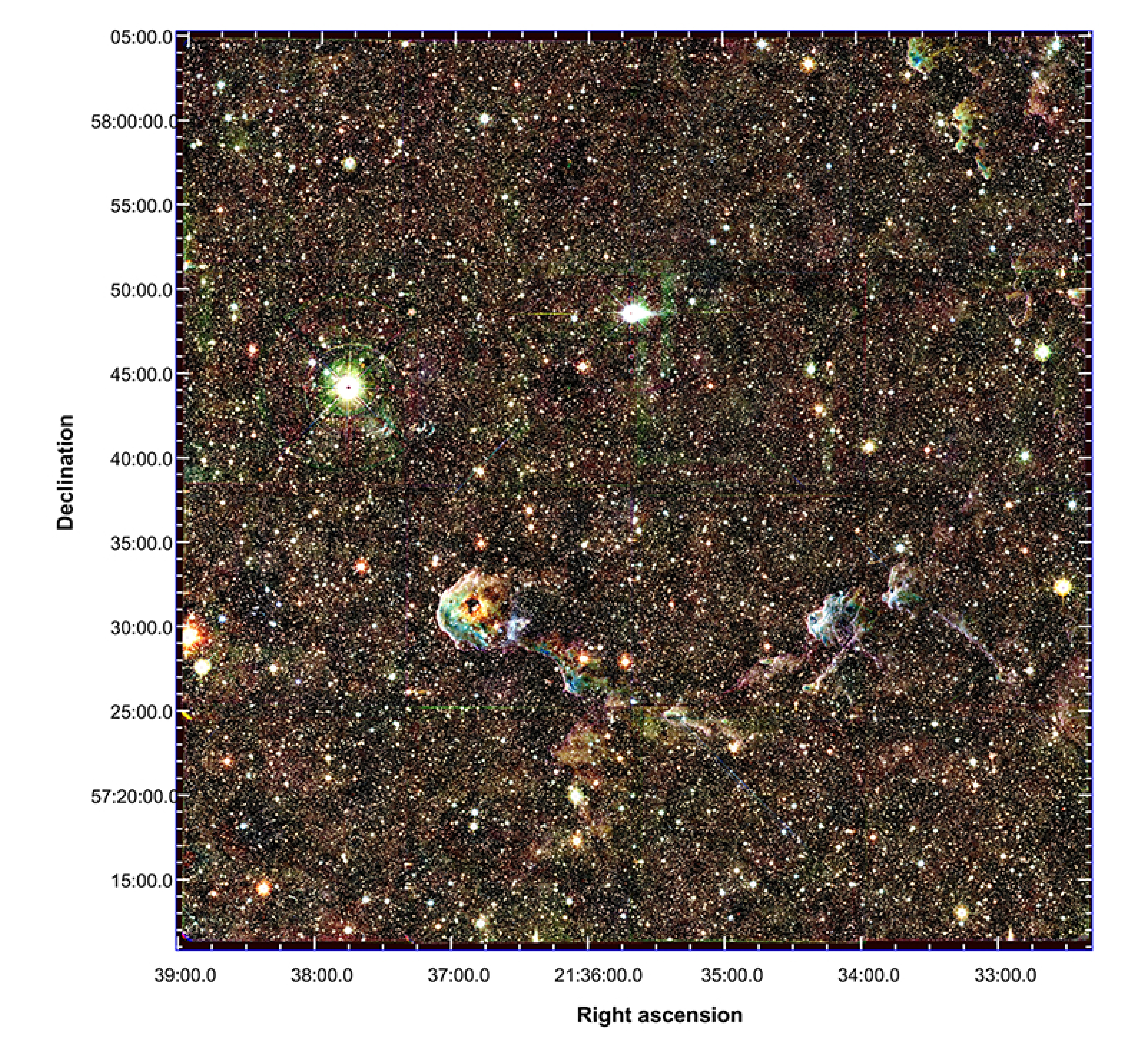}
\caption{A color composite image of TR 37  using UKIRT/WFCAM (blue: $J$, green: $H$, red: $K$). 
North is up and east is to the left.  The coverage was chosen to avoid the bright O6.5+O9 star system 
HD 206267 (which is located just off the field of view toward the left illuminating the IC~1396A globule) 
and to focus on the less studied region to the northwest of the nebula head. 
\label{fig:fig1}}
\end{figure*}

We have built on this previous work by conducting a long-term (2 year span) variability study of
Tr~37 (53\arcmin$\times$53\arcmin~field of view) in the near-IR ($J,H,K$) 
using UKIRT/WFCAM (Figure~\ref{fig:fig1}).  This study includes the globule
IC~1396A and an additional nearby area (the north-western region from IC 1396A) 
to probe a less-studied part of Tr~37. 
In Section 2,  we describe our observations and the data reduction. In Section 3 we present the results and our analysis of variability, 
including variable source selection criteria. In Section 4 we compare our results with previous variability studies, then discuss (1) the 
variability of bright members in Tr~37, (2) variable low mass members, (3) 
new variable member candidates using a color-magnitude diagram analysis to test whether variable stars are likely to be 
cluster members, and (4) characteristics of the highly varying sources. We show that although it has been thought that Tr 37 and IC 348 YSOs are similar in age, the $K$-band luminosity function and $K$ vs. $I-K$ isochrone of Tr 37 are shifted 
brighter by $\sim$ 0.3 $-$ 0.5 mag, indicating that the stars in Tr 37 are younger than those in IC 348. We then introduce simple methods to distinguish different causes of variations: (1) accretion hot spots cause 
larger changes at $J$ than at $K$, and significant scatter in colors, both of which can be identified in appropriate color-color and color-magnitude diagrams; (2) extinction variations have a distinctive track on such diagrams, with little scatter around it; and (3) sources with larger variations at $K$ due to sporadic illumination of disk dust tend to fall on or below the CTTS locus on $J-H$, $H-K$ diagrams. We conclude that: (a) the wavelength dependence of the extinction from the circumstellar disks is noticeably affected by grain growth in the disks; (b) four sources 
show extreme variability and may be related to EXors; (c) the sources that vary predominantly at $K$ nearly all show a color temperature 
near 2000K for the variable component, suggesting the variations arise from dust exposed by turbulence at the inner edge of the circumstellar disk; and 
(d) at least one source appears to have been obscured for an extended period by an optically thick disk component.   A summary and our conclusions
follow in section 5.

\section{Observations and Data Reduction}

\subsection{Observations}

Observations were conducted in $J,H,K$ bands with the Wide Field Camera (WFCAM) \citep{casali2007} on the United Kingdom Infrared Telescope (UKIRT) on Mauna Kea, Hawaii. The focal plane hosts four Rockwell Hawaii-II HgCdTe detectors laid out in a $2\times2$ array. Each detector has $2048 \times 2048$ pixels with a pixel scale of 0.4\arcsec\ pixel$^{-1}$, covering 13.65\arcmin\ on one side. Adjacent detector arrays are separated by a gap 12.8\arcmin\ in width. To fill the gaps and maximize the sky coverage, we dithered the telescope at four positions with a $13.2\arcmin \times 13.2\arcmin$ square pattern, so the detector gaps in one exposure were covered in the previous/next exposure with narrow strips of overlap, yielding the image displayed in Figure~\ref{fig:fig1}. The total sky coverage was a $53\arcmin \times 53\arcmin$ square, ranging from 323.082\degr\ to 324.752\degr ~in RA and from 57.196\degr\ to 58.080\degr ~in Dec. The integration time of each exposure was 10 s, 5 s, and 5 s in $J$, $H$, and $K$, respectively. Each run consisted of 20 consecutive exposures in each waveband, i.e., total integrations of 200, 100, and 100 s respectively. The instrumentation and observing strategy were similar to those used in a number of previous studies of young star variability \citep[e.g.,][]{rice2012, wolk2013b,rice2015}.


The same observational design was executed in all three bands at each epoch on the same field to ensure data homogeneity across time. From July 18, 2014, to July 12, 2016, we have 21 epochs in total\footnote{Not including incomplete observations due to bad weather conditions.} over a time baseline of 725 days, or almost 2 years. The observation start time of each epoch is given in Table~\ref{obs}. Some epochs have more than one run in some bands, as indicated in the notes in Table~\ref{obs}. The seeing ranged from 0.6\arcsec\ at best to 1.4\arcsec\ at worst, with an average of 0.79\arcsec. Although our cadence does not allow study of periodicity, the number of epochs is sufficient to identify the great majority of variable stars \citep{rice2012, rice2015}. 

\begin{deluxetable}{ccc}
\tabletypesize{\tiny}
\tablecaption{Observation Log
\label{obs}}
\tablehead{
\colhead{Epoch Order} & \colhead{Start Time (UTC)} & \colhead{Note}
}
\startdata
1 & 2014 Jul 18, 09:17:53 & \\
2 & 2014 Jul 25, 12:36:29 & \\
3 & 2014 Oct 06, 07:35:49 & \\
4 & 2014 Oct 30, 04:32:44 & \\
5 & 2014 Nov 06, 04:43:15 & \\
6 & 2014 Nov 30, 05:39:27 & \\
\nodata & 2015 Aug 05, 07:59:21 & only first 2 positions in $J$ \\
7 & 2015 Aug 07, 08:24:26 & 1 more run in $J$ \\
8 & 2015 Aug 28, 08:08:53 & \\
9 & 2016 May 24, 13:22:46 & \\
10 & 2016 May 31, 12:21:16 & \\
11 & 2016 Jun 02, 13:34:59 & \\
\nodata & 2016 Jun 07, 13:54:34 & only in $J$ \\
12 & 2016 Jun 09, 13:26:30 & \\
13 & 2016 Jun 10, 13:20:21 & \\
14 & 2016 Jun 20, 11:50:20 & \\
15 & 2016 Jun 23, 11:37:01 & \\
16 & 2016 Jun 25, 13:25:27 & \\
17 & 2016 Jun 26, 12:27:38 & \\
18 & 2016 Jul 05, 11:09:59 & \\
19 & 2016 Jul 08, 10:46:52 & \\
20 & 2016 Jul 10, 10:34:21 & \\
21 & 2016 Jul 12, 12:07:23 & 2 runs in each band \\
\enddata
\end{deluxetable}

\subsection{Data Reduction}

The science images were prepared by the UKIRT/WFCAM pipeline at the Cambridge Astronomical Survey Unit (CASU) \citep{irwin2004, hodgkin2009}. They were stored in the WFCAM Science Archive\footnote{\url{http://wsa.roe.ac.uk/}}  \citep{hambly2008}, where they were also checked for quality and organized for various retrieval strategies. The pipeline includes dividing the physical pixels into 0.2$''$ virtual ones, plus bias, flat field, and dark field corrections. After obtaining the astrometric information \citep{budavari2010}, automated detections and measurements are conducted on the science images to the 2MASS calibration standard \citep{hewett2006,hodgkin2009}. This includes stellar astrometry in equatorial coordinates, aperture photometry with a set of aperture sizes, photometric calibrations, and source classifications in terms of the probability of the source being a star (point-like) or galaxy (extended), or contaminated due to noise or saturated pixels.  The overall classification of a source is obtained by combining independent probabilities of source classifications from individual-epoch images using Bayesian classification rules \citep{hambly2008}. A ``stellar'' source has to have an overall probability of 90\% or greater to be considered point-like. The detection tables of individual images were then merged to cross-match the detections of the same sources in different wavebands and at different epochs. The outcome is a table of all sources detected in the field with astrometry and $JHK$ photometry, both time-averaged and by individual epochs, as well as source classifications. When a single time-averaged measurement of a star is needed, we use the astrometry and photometry based on the stacked mosaic image in the waveband of interest. 

The pipeline-reduced measurements are available in the database WSERV9v20170222, from which we retrieved them via SQL inquiry. WSERV9 is a combined program made up of U/14B/UA16, U/15B/UA20 and U/16A/UA19 and the data were processed as a correlated multi-epoch project to produce contemporary colours in $JHK$, much like the WFCAMCAL data set described in \citet{cross2009} or the Orion Nebula Cluster \citep{rice2015}. The data can be made available by contacting Serena Kim (serena@as.arizona.edu).

\subsection{Other data sets}

We make use of a number of other data sets, as described in the Appendix.

\subsection{The Extensive Catalog}

In this work we are only interested in stellar sources with accurate photometry. To allow for systematic as well as 
statistical errors, we adopt 0.05 magnitude (signal-to-noise ratio $S/N \gtrsim 20$) in the stacked all-epoch mosaic images in all three bands as the threshold of photometric uncertainty for inclusion in our study. This threshold rejects bright stars by the requirement of non-saturation and rejects faint stars by the maximum photometric error. A total of 111,657 qualified stars is found in the entire UKIRT/WFCAM field of view. We describe this list as the ``extensive catalog'' representing the largest set of stars with possibly useful measurements, which is not selected based on individual-image measurements or any time-domain information. The magnitude ranges in the extensive catalog are from 10.6 to 20.5 in $J$, 10.3 to 19.6 in $H$, and 9.9 to 19.3 in $K$. As a sample of the data products, Figure~\ref{fig:fig1} shows the mosaic image of the Tr 37 in false color using deep stacked UKIRT/WFCAM images.  

\section{Selection of Stars with High Quality Data}\label{selection}

Not every star in the extensive catalog is suitable for variability analysis. Many of them have post-processing error flags from the pipeline at some epochs in some wavebands. These are warnings that the measurements in question may be biased or have other potentially significant data issues \citep{hodgkin2009}. After visually inspecting many of these flagged stellar images, we find that the flags can arise from quite a few factors related to the observational conditions, such as close binaries resolved only during the best seeing conditions, slight focus inaccuracy, or high sky brightness at some epochs. Other than the flagged observations, some stars are simply undetected at some epochs, possibly because of unfavorable observational conditions. We assess the quality of individual photometric measurements by classifying a star at each epoch to either have a good observation, an error-flagged observation, or a missing observation. Only good observations are used to calculate the variability classifications.

\begin{figure}
\epsscale{1.0}
\plotone{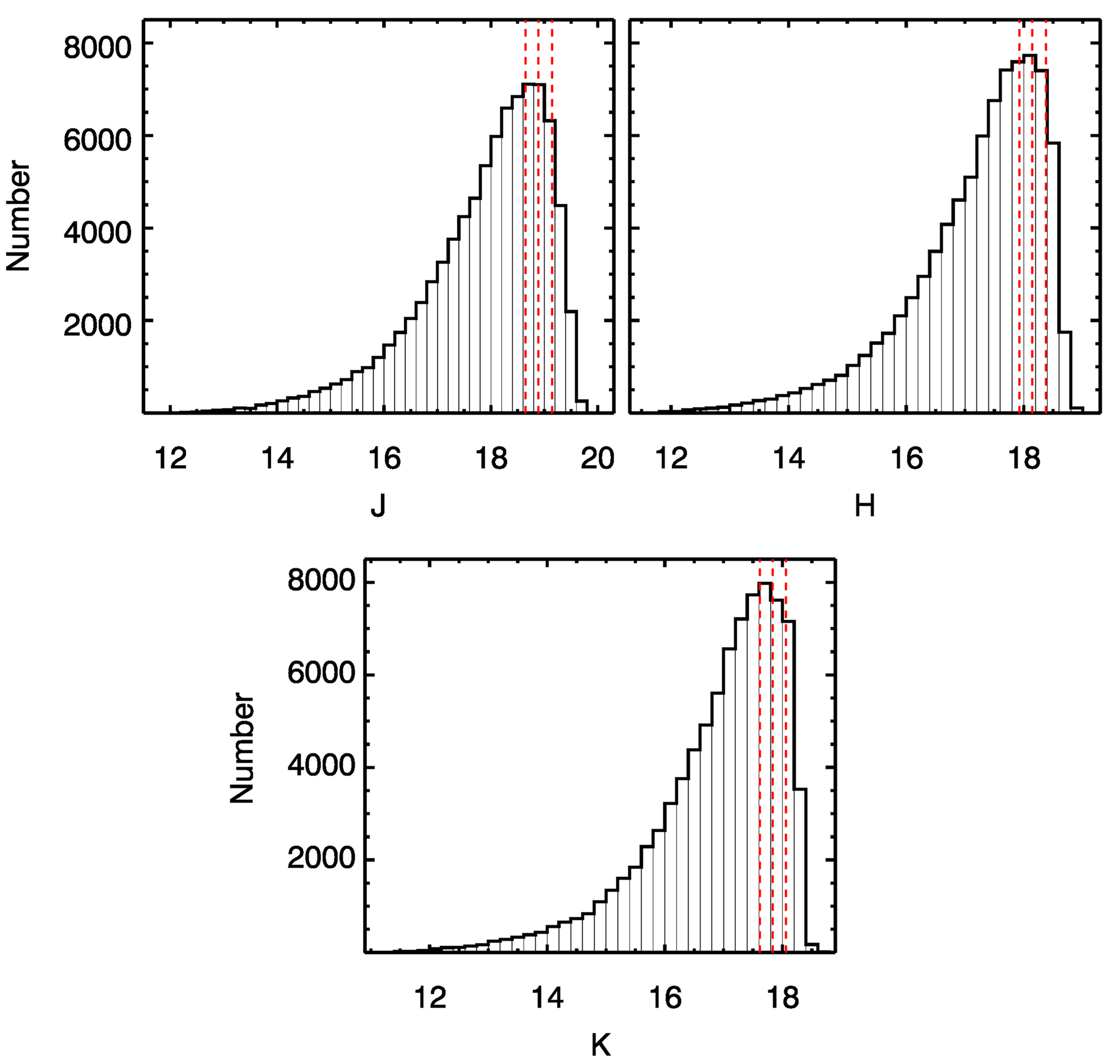}
\caption{Histograms of the magnitude distribution of all detected stars in the $JHK$ band variability catalog. In each plot, the red dashed lines, from right to left, label the brightness levels for 70\%, 80\%, and 90\% of all detected stars. 
\label{hist}}
\end{figure}

Given that each star is covered by at least 24 images in $J$ and 21 observations in $H$ and $K$\footnote{Stars that lie in the narrow strips with overlapped coverage at consecutive dithers may be observed by a multiple of 24/21 times.} (Table~\ref{obs}), we require a star to have at least 18 good observations in each of the three bands for its variability classification to be considered reliable. This threshold was set after examining sample light curves for sources with fewer good observations; 18 observations eliminated virtually all questionable cases. Additional tests of the method included plotting the possible variables as an image and adjusting the procedures to eliminate any suspicious spatial clustering (e.g., along the ``seams'' in our coverage at the extreme edges of the arrays). A total of 85,814 stars qualify, which make up our final catalog for the purpose of variability analysis, i.e. the ``variability catalog,'' which is a subset of the ``extensive catalog.'' The magnitude ranges in the final variability catalog are from 11.7 to 20.1 in $J$, 11.5 to 19.2 in $H$, and 11.2 to 18.7 in $K$. The magnitude histograms in each band are shown in Figure~\ref{hist}; the behavior suggests completeness limits of $\sim$ 18.3, 17.7, and 17.4 respectively at $J$, $H$, and $K$. In this section, we describe the analyses conducted on this catalog to identify variable stars.

 We assign one of five quality flags to each star in the extensive catalog as described in Table~\ref{Q_flag}. The variability catalog only includes stars with a quality flag $\geq1$.  It is likely that some stars with less than 18 good observations in at least one band (which thus are excluded from the variability catalog) may still be true variable stars. Although some of these stars are probably genuinely variable, many of these cases may have other causes such as asteroids or artifacts from bright stars. We do not consider their variability classifications further.

\begin{deluxetable*}{ccl}
\tabletypesize{\scriptsize}
\tablecaption{Definition of the Quality Flags for Time-domain Measurements
\label{Q_flag}}
\tablehead{
\colhead{Quality} & \colhead{VarCat\tablenotemark{a}} & \colhead{Criteria}
}
\startdata
2 & Y & No post-processing errors in any band at any epoch\\
  &   & (at least 21 ``good'' measurements in any band).\\
1 & Y & At least 18 ``good'' measurements (without post-\\
  &   & -processing errors) in each of the $JHK$ bands.\\
0 & N & Less than 18 ``good'' measurements \\
   &   &  in at least one band. \\
-1 & N & No reliable single-epoch photometry \\
  &  &  in at least one band.\\
-2 & N & No minimum/maximum magnitude in at least one  \\ 
  &   &  band (one or no good single-epoch photometry).
\enddata
\tablenotetext{a}{``Y'' if included in the variability catalog, ``N'' if not included.}
\end{deluxetable*}

\section{Analysis of Variability}\label{analysis}

This section describes how we have combined two approaches to identifying variable stars from our ``variability catalog'': (1) analysis of the rms fluctations in the measurements of the stars; and (2) application of the Stetson Index that evaluates correlated variations in multiple photometric bands. Application of these two methods together leads to identification of 289 variable stars with K magnitude $<$ 17, a sample that we analyze in the next section. 

\subsection{Initial Variability Identification Through RMS Fluctuations}

For each of the $J,H,K$ bands, a set of statistics is calculated for each star in the catalog, including the mean, median, rms, minimum and maximum magnitudes, and skewness \citep{sesar2007, cross2009}, based on the photometry with the best aperture. The best aperture is defined as the aperture size that gives the lowest rms for the photometry of the object with the appropriate aperture corrections, an approach that yields better measurements in crowded regions than using a single, fixed aperture \citep{irwin2007, cross2009}. The means and rms's for all sources are used to derive a fit to the expected scatter as a function of magnitude \citep[e.g.,][]{strateva2001}, which we call the ``expected rms'' or $xExpRms$\footnote{We use the notation supplied with the pipeline products to expedite comparison with other UKIRT programs.}. This gives the photometric uncertainty for band $x$ ($J$, $H$, or $K$) based on an individual image. For a non-variable star, this error should be an accurate estimate of the real scatter of the photometry in a time series, and is similar in principle to the error estimation in \citet{hodgkin2009} as applied by \citet{rice2012, rice2015}. 

Variability is identified as in \citet{cross2009}, using attributes from the database. That is, the ``expected rms'' is subtracted, in quadrature, from the real rms scatter of the photometry at different epochs ($xMagRms$). This reveals the variability of the star, in the form of ``intrinsic rms ($xIntRms$),'' beyond the noise in the system. Specifically, in $x$ band where $x$ equals either $J$, $H$, or $K$

\begin{eqnarray}
xMagRms^2 & = & \frac{1}{xndof} \displaystyle\sum_{i=1}^{xndof} \left( x_{i} - xmeanMag \right)^2\\
xIntRms & = & \left( xMagRms^2 - xExpRms^2 \right)^{1/2}
\end{eqnarray}

\noindent
where $xndof$ is the number of good observations and $xIntRms$ is the rms intrinsic variability, both defined in $x$ band. To identify a star as a candidate for variability, we require that the probability 
of the source being variable be $>$ 0.9 in each band, estimated by integrating the reduced $\chi^2$ function:

\begin{footnotesize}

\begin{equation}
xchiSqpd = \frac{1}{xndof - 1} \chi_{x}^2 = \frac{1}{xndof - 1} \displaystyle\sum_{i=1}^{xndof} \left( \frac{x_i - xmeanMag}{xExpRms} \right)^2
\end{equation}

\end{footnotesize}

\noindent
$xchiSqpd$ is calculated under the null assumption of non-variability with the expected rms calculated above; for details see \citet{cross2009}.

We then conduct a second test on the candidate variables. We identify a star as a confirmed variable only if the weighted average ratio of the intrinsic to expected rms over all bands is greater than 3:

\begin{equation}\label{overall_def}
wsrms = \frac{\displaystyle\sum_{x=J,H,K} w_x \frac{xIntRms}{xExpRms}}{\displaystyle\sum_{x=J,H,K} w_x} \geq 3
\end{equation}

\noindent
where the weight $w_X$ is proportional to the number of good observations in each band; the band with the most observations has a weight $w_X = 1$. 

There are 328 stars that pass both of these tests (we designate variables by {\sc VarClass} = 1, with the remainder as  {\sc VarClass} = 0). 

\subsection{Application of the Stetson Index}

We used the Stetson Index, $S$, \citep{stetson1996} as an alternative way to identify variable stars:

\begin{equation}\label{overall_def}
S = \frac{\displaystyle\sum_{i=1}^{p} g_i ~sgn\left(P_i\right) \sqrt{|P_i|} }{\displaystyle\sum_{i=1}^p g_i}.
\end{equation}

\noindent
Here, $p$ is the number of pairs of simultaneous observations of a star, $g_i$ is the weight of the $i^{th}$ measurement, and $P_i$ is the product of the normalized residuals of 
the two observations where the normalized residual of the $i^{th}$ observation is

\begin{equation}
\delta_i = \sqrt{\frac{n}{n-1}} \frac{m_i - \overline{m}}{\sigma_i},
\end{equation}

\noindent
where $n$ is the number of observations in a band, $m_i$ is the magnitude in the $i^{th}$ observation, 
$\sigma_i$ is the observational uncertainty in that observation, and $\overline{m}$ is the average magnitude. 

Nearly all of our observations provide simultaneous $J$, $H$, and $K$ measurements of a source. We computed $S$ for each pair of colors, i.e., three values for each source\footnote{We deleted the most discordant of all repeated photometry of a star before the calculating Stetson index if it was made within $\sim$ 50 min from the previous one, roughly the time needed to complete one epoch. This removed double photometry of the stars at the edge of a field where the second telescope pointing resulted in a second measurement.}. We then conducted a number of tests to determine the optimum way to utilize this information. First, considering the three indices separately might be important if there are sources that vary significantly in two adjacent bands but not in the third. However, we found no convincing cases of such behavior from comparing the band-pair Stetson Indices. In addition, as discussed in the next section, there were in the end no examples of single-band variations that escaped notice in other ways. We therefore averaged the three values of $S$ for each source. We found that these averages had a Gaussian distribution; assuming symmetry (i.e., that the number of false variability identifications can be deduced from the low side of the Gaussian), we found that $S \ge 1$ is required to avoid false identifications fully (\citet{rice2012, rice2015}, for example, adopted the same threshold). Since we do not know that the Gaussian is a perfect fit, we adopt this conservative criterion to identify variable stars purely through the Stetson index. 

We adopted a different approach for sources where our $\chi^2$ analysis indicated variations. Already the gaussian distribution indicates that $S \ge 0.7$ indicates a probability $<$ 0.1\% of a false identification of variability. We examined the light curves of sources with $S > 0.5$ and found that those with $S < 0.7$ frequently did not present convincing evidence for variations, but above this value most cases did imply variations, with the number of false identifications rapidly decreasing as $S$ approached 1. Therefore, if a source was identified as a variable in the $\chi^2$ analysis {\it and} had $S \ge 0.7$ {\it and} also passed visual inspection as showing true variations, we accepted it as a true variable.

\begin{figure}
\epsscale{1.2}
\plotone{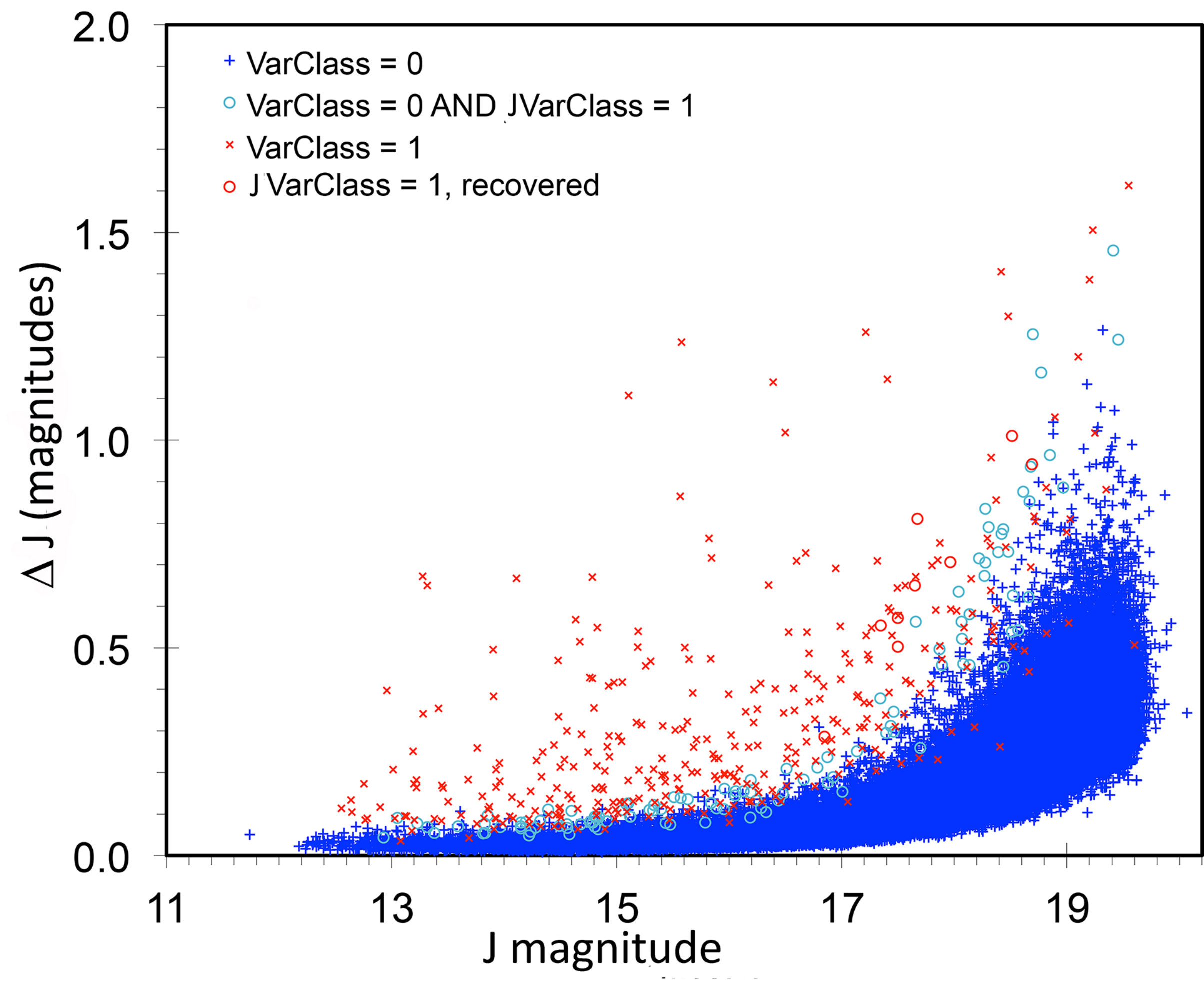}

\caption{Peak-to-peak amplitude in magnitudes as a function of the mean $J$ magnitude in the mosaic image. The points are colored to show the overall variables ({\sc VarClass} = 1) as red crosses. Single-band variables that do not qualify as overall variables ({\sc JVarClass} = 1 but {\sc VarClass} = 0) are marked with cyan circles, unless they have $xchiSqpd \geq 20$ in that band, in which case they are plotted as red circles. Other non-overall variables are in blue. Stars with peak-to-peak amplitude higher than 2 mag are not shown. The stars indicated in red circles are candidates to add to the variable category; they are clearly differentiated from the distribution of non-variable stars (blue).
\label{var-amp}}
\end{figure}

\subsection {Single-Band Variability}

Both the $\chi^2$ and Stetson Index approaches emphasize simultaneous variability in more than one spectral band. We carried out a third search in case there are variables that exceed the threshold for detection only in a  single band and that were missed by these approaches. For example, a non-variable star with a highly dynamic dusty disk might show significant variability in $K$ band because of the varying excess, but in $J$ band where no significant excess is present, the star itself may not have detectable variations. Such a star might be disqualified by the all-band requirement. 

Without the confirmation provided by variations in multiple bands, we need to be alert to possible spurious causes of variations. An example is a relatively low ratio of signal to noise. To probe this possibility, in each band we calculate the average and median flux densities, measured on the mosaic images in each band, of all identified variable stars in that band ({\sc xVarClass} = 1, where {\sc x} designates  the band, $J$, $H$, or $K$) and compare them with those of the subset in which the stars are non-overall variables ({\sc VarClass} = 0).  Although many of the single band variable candidates are faint, suggesting low signal to noise, there are a few that are bright enough that the variations might be real, as  illustrated for the $J$ band in Figure~\ref{var-amp}.  We use the reduced $\chi_{red}^2$, or $xchiSqpd$ (equation (3)), in a more stringent test to identify single-band variations. 
Given the possibility of false identifications, we have been conservative. A star is considered to vary if it is identified in any of the $J,H,K$ bands with $\chi_{red}^2 \geq 20$ relative to the nominal measurement errors (given systematic noise of $\sim$ 2\%, the significance is typically $3-5 \sigma$). This recovers 36 stars in total (11 stars by $\chi_{red}^2$ in $J$, 6 in $H$, and 19 in $K$). However a review of the individual $JH$, $HK$, and $JK$ Stetson Indices and of the  light curves of these objects led to the conclusion that no high-confidence variables had been added to our sample. 

\subsection{Summary}

In total, there are 359 stars with detected variations in our final ``variability catalog'', an incidence of 359/85814 = $0.42 \pm 0.02\%$  covering the full  survey field. We will show that 1/3 of them are probable cluster members, implying that the identified variables among true field stars are $\le$ 0.3\%. This value can be compared with the detection of variability in $\sim$ 1.6\% of the field stars with similar data by \citet{wolk2013a} in a field at the same Galactic latitude and just 7 degrees away in Galactic longitude. The greater number of epochs in their monitoring might account in part for the higher detection rate, but with only four nights of observation \citet{pietrukowicz2009} detected variability in 0.7\% of field stars in Carina.  The comparison indicates that our analysis is conservative and does not return a large number of false indications of variability. The lack of field star contamination is important for the analysis we will conduct on the sample in the following section (Section~\ref{sec:analysis}).

Figure~\ref{var-amp} and similar statistics for $H$ and $K$ show where the distribution of variability amplitudes will begin to be biased against small amplitudes at faint limits. Analyzing this effect at $K$ as the fiducial band for our study, the survey should be essentially unbiased in variability amplitude down to $K$ $\sim$ 16, will miss about the 20\% of the smaller amplitude sources at 16.5, and will become increasingly incomplete in the smaller amplitude range below 17 mag. In addition, at faint limits it becomes increasingly difficult to identify faint nearby sources that might be responsible for false indications of variations due to the effects of seeing variations from night to night, and interference by the wings of the PSFs of nearby bright sources also becomes more likely. We therefore base the analysis of the variable source population only on sources with $K < 17$. Based on the color-magnitude diagram in Section~\ref{sec:CMD}, we find 121 possible cluster members among the variable sources. The Gaia DR2 parallaxes of two of these sources indicate they are foreground to the cluster. In addition,  63 of the apparently variable stars that appear to be non-members are fainter than K = 17. That is, the study is based on 119 variable probable cluster members with $K < 17$, listed in Table 3, that we will distinguish from 170 variable nonmembers based on placement on the CMD (Section~\ref{sec:CMD}).

\begin{figure*}
\epsscale{1}

\plotone{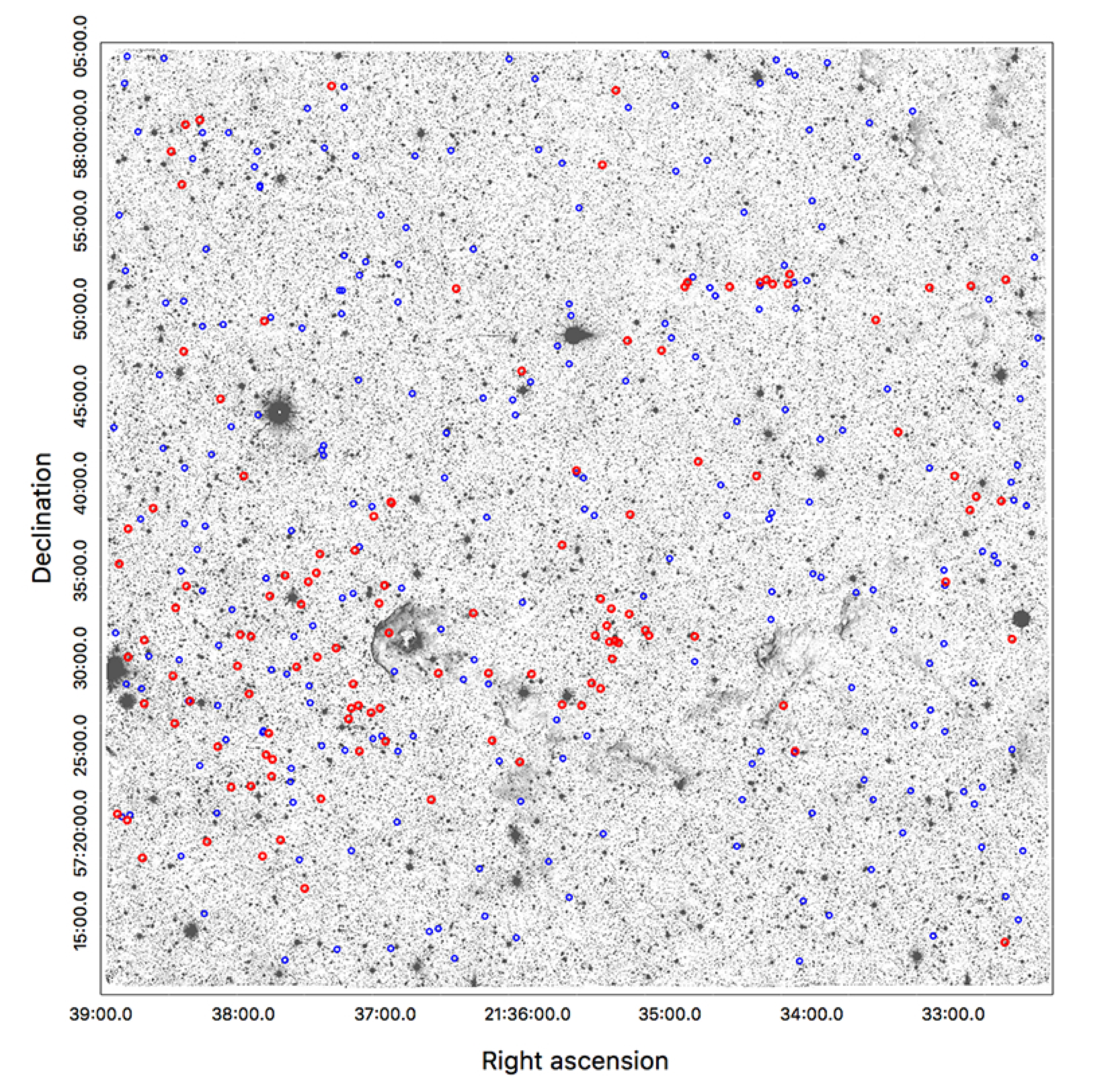}
\caption{A UKIRT/WFCAM image of the surveyed region in $K$ band. North is up and east is to the left; 
Variable cluster members are marked by magenta circles 15\arcsec\ in radius. They tend to 
lie in the vicinity of IC 1396A. Small blue circles 8\arcsec\ in radius mark additional variable stars; these stars are more uniformly spread over the field.
\label{2Vars}}
\end{figure*}

Table 4 lists the variable stars that are nonmembers according to the CMD analysis; we include those fainter than $K$ = 17 but include them only as possible variables. Figure~\ref{2Vars} shows the locations of the variable stars brighter than K = 17 mag, differentiating probable cluster members from those that are probably in the field.

\section{Analysis}
\label{sec:analysis}

The Cep OB2 region, including IC 1396A and the surrounding clusters NGC 7160 and Tr 37, has been previously observed for YSO variability at other wavelengths. In the first subsection of this discussion, we compare known variable stars (from two studies, one in the $R$ and $I$ and the other in the {\it Spitzer} IRAC photometric bands) with our variability catalog.  In the following subsection, we discuss the variability of the relatively bright Tr 37 cluster members identified by \citet{sicilia-aguilar2004, sicilia-aguilar2005, sicilia-aguilar2006b}. In the next subsection, we carry out a similar study of the fainter sources from a different region of the cluster identified by \citet{sicilia-aguilar2013}. These subsections demonstrate that we detect variations in a large fraction ($65 \pm 8$\%) of the {\it bona fide} cluster members. Therefore, in the following section we identify a homogeneous sample of cluster members on the basis of variability and a location on the $K$, $I-K$ color-magitude diagram that is consistent with cluster membership. This sample includes many newly identified members, in particular, faint ones. In the final subsections, we discuss the characteristics of the variability of individual cluster members from this sample that are of particular interest. 

\subsection{Previous Studies of Variability}

In their Tables 6 and 7, \citet{sicilia-aguilar2004} gave variability identifications of stars in the region in the $R$ and $I$ bands. The two tables combined have 77 stars in total, 31 of which are within our UKIRT/WFCAM field of view and not saturated in any band. We cross match the 31 stars and find that 30 of them have counterparts within 2\arcsec ~detected by UKIRT. The maximum matching distance is 1.2\arcsec.\footnote{There is one star, 12-583 in Table 6 \citep{sicilia-aguilar2004}, that has no match in our catalog within 3\arcsec. There is a source detected 0.88\arcsec\ away, but it is classified as a galaxy (extended source) with very high confidence ($P_{galaxy} = 0.999657$) by the pipeline \citep{hambly2008} and thus is excluded from our catalog. Nonetheless, \citet{sicilia-aguilar2010} indicate it is a star of spectral type M0; an inspection of the stacked image finds a tail of the source significantly extended to the east. Though a binary star is one of the possibilities, for data homogeneity we do not count any extended source identified by the WSA pipeline.}
Among the 30 stars with stellar counterparts in our catalog, only nine probable members are in the variability catalog. Six of these nine are classified as overall variables: 11-2037, 11-2131, 14-1017, 11-383, 13-924, and 12-1091, in the nomenclature of  \citet{sicilia-aguilar2004}. (The 3 stars that we do not find variable in $JHK$ are 12-1081, 12-1825, and 13-1161.) These stars are listed in Table 5 along with additional information about the variability of the bright cluster members.

We now turn to the mid-infrared. \citet{morales-calderon2009} found 41 stars to be variable out of a sample of 69 in the region in the {\it Spitzer} IRAC bands from 3.6 to 8.0 \micron, with 40\% of the sample showing peak to peak amplitudes $>$ 0.1 mag.  To explore this result, we evaluated the variations of the 41 stars listed in \citet{morales-calderon2009}, Table 3. Although our observations are removed in time from those reported in this table, we can look at the type of the {\it Spitzer}-detected variability to compare with our results. Twenty five of the stars are within the parameters of our study - this excludes stars that are above our saturation limits in $JHK$ and those that are so red that we would not have reliable $I$ measurements and hence where we could not determine them to be cluster members based on the $K, I-K$ color magnitude diagram (see Section 5.4). The distribution of peak-to-peak amplitudes in our study implies that we are relatively incomplete for amplitudes $<$ 0.1 - 0.12 mag, corresponding to RMS variations of $\sim$ 0.03 mag\footnote{Making the usual assumption that the peak-to-peak variation is four times the RMS one.}. Assuming that the $JHK$ variations should have similar sizes as those in the IRAC bands, we eliminate stars below this variability threshold, leaving 19. 
We have found variations in only five of these stars, i.e., 26\%, of which only four are identified as variable by \citet{morales-calderon2009}. Although nominally consistent with the 40\% variability above 0.1 mag reported by \citet{morales-calderon2009}, the comparison is surprising in the low rate of sources with variations seen in both wavelength ranges. 

\begin{figure}
\epsscale{1.2}
\plotone{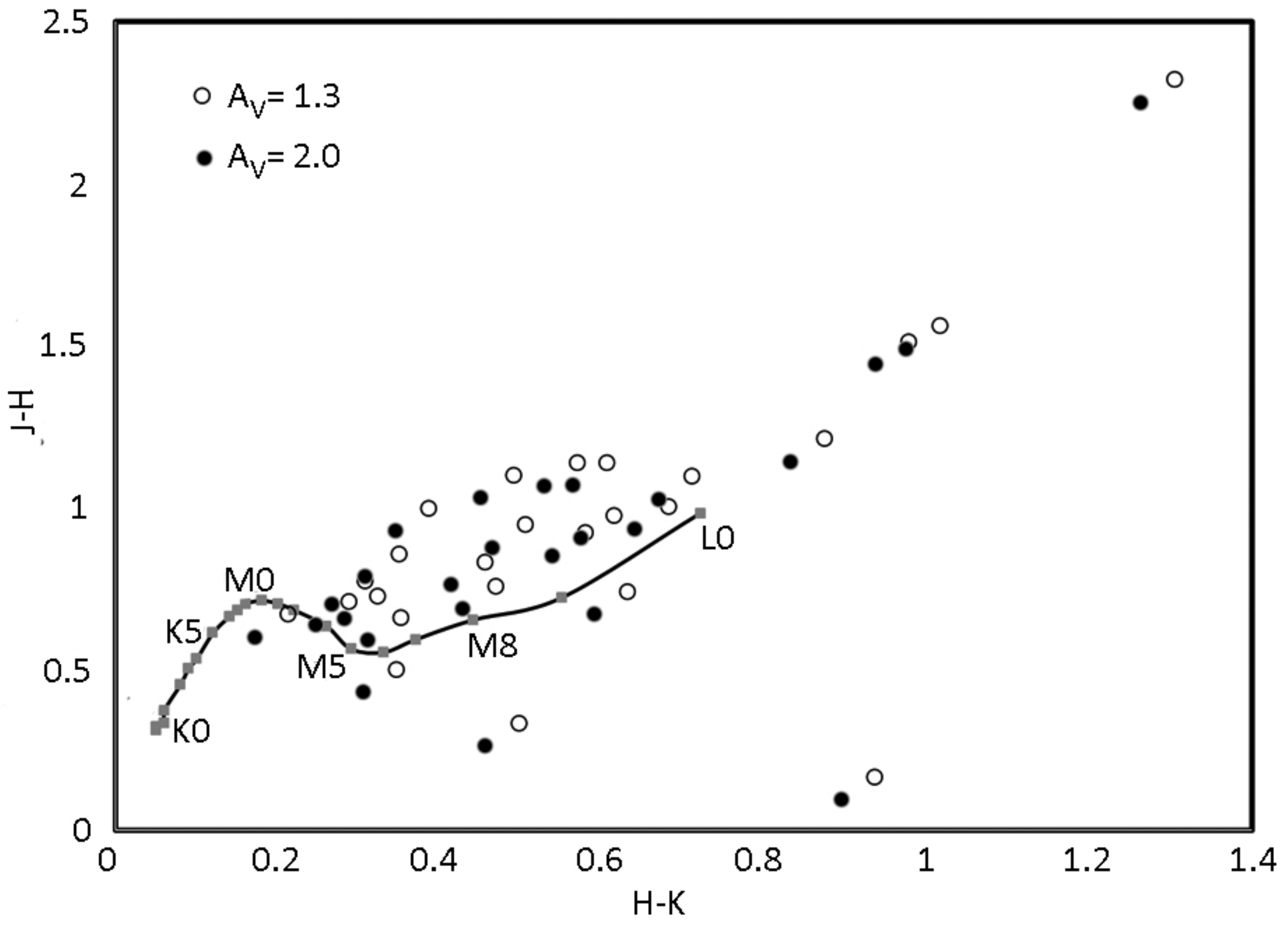}
\caption{Placement of stars variable in the IRAC [3.6] and [4.5] bands \citep{morales-calderon2009} on the $J-H$, $H-K$ color-color diagram. The locus of young low-mass stars and brown dwarfs is also shown, taken from \citet{luhman2010}. The filled dots are for $A_V = 2$ and the open ones for $A_V = 1.3$.
\label{trans}}
\end{figure}

One explanation is that the emission mechanism in the IRAC bands is different from that in $JHK$, e.g., dominated by the photospheric variability, extinction changes, and accretion events onto the stellar surfaces at $JHK$ and by circumstellar disk behavior in the IRAC bands. For example, two of 
the four variables from both studies (i.e., 182 \& 192) appear to have $JHK$ variability due to extinction variations (see Section 5.4). However, for the first the rms variations are the same size at [4.5] as at [3.6], which is not expected if the variations in these bands are also due to extinction; i.e., the  variations seen by  \citet{morales-calderon2009} are not of the same origin as those we find at $JHK$.   Figure~\ref{trans} examines this possibility more generally. The figure tests how many of the IRAC variables may {\it not} have excesses in $JHK$. It shows the location on the $J-H$ vs. $H-K$ diagram of the IRAC variables, corrected for two possible levels of extinction, $A_V$ =1.3 and 2, covering the range of plausible values for cluster members. These points are compared with the young stellar locus from \citet{luhman2010}. Six of the stars have colors compatible with stars of type $\le$ M6 with no excess emission (one of these has rms variations $<$ 0.03 mag); two more have colors consistent with stars between M6 and M8 in type. At the same time, all of the sources have [3.6], [4.5], [5.8], and [8.0] colors consistent with Class II sources \citep{gutermuth2009}. That is, a number of these sources indeed are likely to be dominated at $JHK$ by photospheric emission and to have variable infrared excesses only in the IRAC bands.  

Another possibility is that the lack of common variability arises in part because of the time interval between the observations; the {\it Spitzer} data used by \citet{morales-calderon2009} were obtained between January 24 and February 6, 2008, 7 to 9 years prior to our observations, and young star variability can be episodic over such time intervals \citep{audard2014}. 

\subsection{Variability of Additional Members of Trumpler 37 from \citet{sicilia-aguilar2004, sicilia-aguilar2006b}}

A series of papers led by Sicilia-Aguilar \citep{sicilia-aguilar2004, sicilia-aguilar2006b, sicilia-aguilar2010, sicilia-aguilar2013, sicilia-aguilar2015} has used a combination of {\it Spitzer}, {\it Herschel}, and 2MASS infrared data, high resolution spectroscopy with {\it Hectochelle} on the MMT, and optical photometry to characterize about 160 sources in the Tr 37 field. Our survey for variability includes about 60\% of this list, i.e., the 106 sources listed in Table 5. As discussed below, we gathered variability information for these sources in the $r$ and $i$ bands, from 2MASS at $JHK$ in addition to our $JHK$ survey, and from the ALLWISE $W1$ and $W2$ band measurements. 

We have determined $r$ and $i$ variability from the IPHAS survey \citep{barentsen2014} on MJD 53,245, Pan-STARRS \citep{chambers2016} around MJD 55,990, and \citet{sicilia-aguilar2004}. All of the measurements from IPHAS and Pan-STARRS were transformed onto a common system using linear fits with the $r - i$ color as the independent variable. We then compared various sets, setting a conservative criterion for identifying that the source had varied: a discrepancy of $\sim$ 0.15 magnitudes (the value varied slightly from one data set to another) in both $r$ and $i$ and in the same direction. In addition, \citet{sicilia-aguilar2004}  identify stars that varied in $R$ and $I$ over the 5 - 7 consecutive nights of data obtained in September, 2000, i.e., around MJD = 51,800. We characterize these variations as fast (F) and those between the different data sets, typically separated by years in time, as slow (S).

For our WFCAM data, we indicate detected variability as $V$ and where none has been detected as $N$.  The WFCAM saturates near magnitude 11.2 (in $K$), so we cannot study variations of the brightest sources with it.

Tr 37 is among the regions scanned at longer integrations as the 2MASS all sky survey \citep{cutri2003} was being completed, producing measurements in the 2MASS 6X data base \citep{cutri2012}. The all sky survey scanned the cluster on October 7, 2000 (MJD 51,824), the 6X survey surveyed 11 days later (MJD 51,835). We searched for variability between these two data sets as described in the Appendix.  We have identified all of these variations as slow, since there is no indication of day-to-day behavior.

We also obtained data on these sources from WISE \citep{wright2010} in the $W1$ (3.4 $\mu$m) and $W2$ (4.6 $\mu$m) bands. There is a rapid cadence of these measurements from MJD 55,373 through 55,375, a few measurements on MJD 55,381, and another rapid series for MJD 55,553 through 55,555. Identification of variability within these data is discussed in the Appendix. Of the 29 sources with valid measurements from WISE and with coverage in our UKIRT monitoring in $JHK$, ten are variable in both spectral regions, a higher fraction than for the {\it Spitzer} monitoring reported by \citet{morales-calderon2009}, despite the poorer sensitivity to variability with WISE.  This result suggests that the low rate of overlap in variability in $JHK$ and the IRAC sample of  \citet{morales-calderon2009} may in part result from low-number statistics.

Table 5 includes additional information to assist with interpreting our results. We have used the ALLWISE $W1 - W2$ color to search for sources with excesses. This color is only slightly affected by extinction, and the excesses tend to be large so we have not attempted to deredden it. Although the color baseline is small, the two colors are measured at the same time - an important consideration since most of the Tr 37 sources are variable. We have also indicated the accretion rates estimated by \citet{sicilia-aguilar2010}. We indicate cluster membership, based on the spectroscopic survey in \citet{sicilia-aguilar2006b}, where we have combined the indicators of membership into a single metric giving priority to radial velocities over the presence of emission lines\footnote{Specifically, we have indicated Y(e) as P, Y(r) as Y, P(r) as P, P(e) as P, Y(e)N(r) as PN, Y(e)P(r) as Y,  Y(e)PN(r) as P:, and P(e)N(R) as PN, where the first designation is from  \citet{sicilia-aguilar2006b}.} This priority is important because of the complexity of the line of sight toward Tr 37 and the likelihood that there are young stars projected onto the cluster that are not associated with it. 

There are 78 stars from Table 5 that are both probable cluster members and have some information on variability. These 78 stars confirm expectations for relations among the different observational parameters. For example, 50 of these stars have measurements of accretion, of which 16 have accretion at a significant level \citep{sicilia-aguilar2010}. All 13 of these 16 stars with adequate variability data are also found to vary in one or more of the three wavelength regions in the table, and all 13 with valid WISE data have excesses above the stellar photospheric color in $W1 - W2$.  Of the stars with variability information in $JHK$, 40/60 or 67\% have been observed to vary. However, one of these stars is a possible foreground object (see footnotes to table) leaving the fraction of cluster members observed to vary at 40/59 = 68\%.  For all 78, about 2/3 have been observed to vary in one band or another, and about 1/2 have $W1-W2$ excesses (of those where we can measure them, omitting ones in confused regions). These numbers imply that variability can produce identifications as complete as those based on infrared excesses, although we have used a relatively large number of epochs and bands and smaller-scale studies of variability will be less complete. 

\subsection{Variability of Lower-Mass Tr 37 Members from \citet{sicilia-aguilar2013}}

\citet{sicilia-aguilar2013} extended their previous work to fainter stars, providing a list of members of Tr 37 down to the mid-M-star spectral type (e.g., covering most objects of stellar but not so faint as to be of brown dwarf mass). Identifying low-mass members of Tr 37 has been challenging because it is projected onto a rich field of stars; there are more than 100,000 stars detected within our UKIRT survey. \citet{sicilia-aguilar2013} addressed this issue by complementing isochrone models and optical color-magnitude diagrams with infrared excesses to select pre-main sequence candidates. They gave priority in their spectroscopic observations to the infrared excess sources, resulting in a possible bias toward such stars in their sample.  Although the initial selection in \citet{sicilia-aguilar2013} sampled M type stars, their confirming spectroscopy (spectra obtained with Hectospec on the Multiple Mirror Telescope (MMT)) 
did not go as deep as very low mass late M stars/brown dwarfs.  
The confirmation rate for the sources with infrared excesses was high ($\sim$ 90\%), but it was much lower (5 - 20 \%) for those without excesses; the sample is therefore largely selected on the basis of having circumstellar disks. Table 6 lists the sources in \citet{sicilia-aguilar2013} with valid variability observations (i.e., in our variability catalog). One of these stars, S13-081, appears to be foreground to the cluster according to its Gaia parallax. Thus, of the cluster members, 19/32, or 59\%, are variable in our data. If we combine this result with those in Sections 5.1 and  5.2, the incidence of sources with detected JHK variability is 65/100 = 65 $\pm$ 8\%. In the following section, we use variability as a means to identify additional faint cluster members. 

\subsection{New Variable Member Candidates - Color-Magnitude Diagram Analysis}
 \label{sec:CMD}

\begin{figure}
\epsscale{1.2}
\plotone{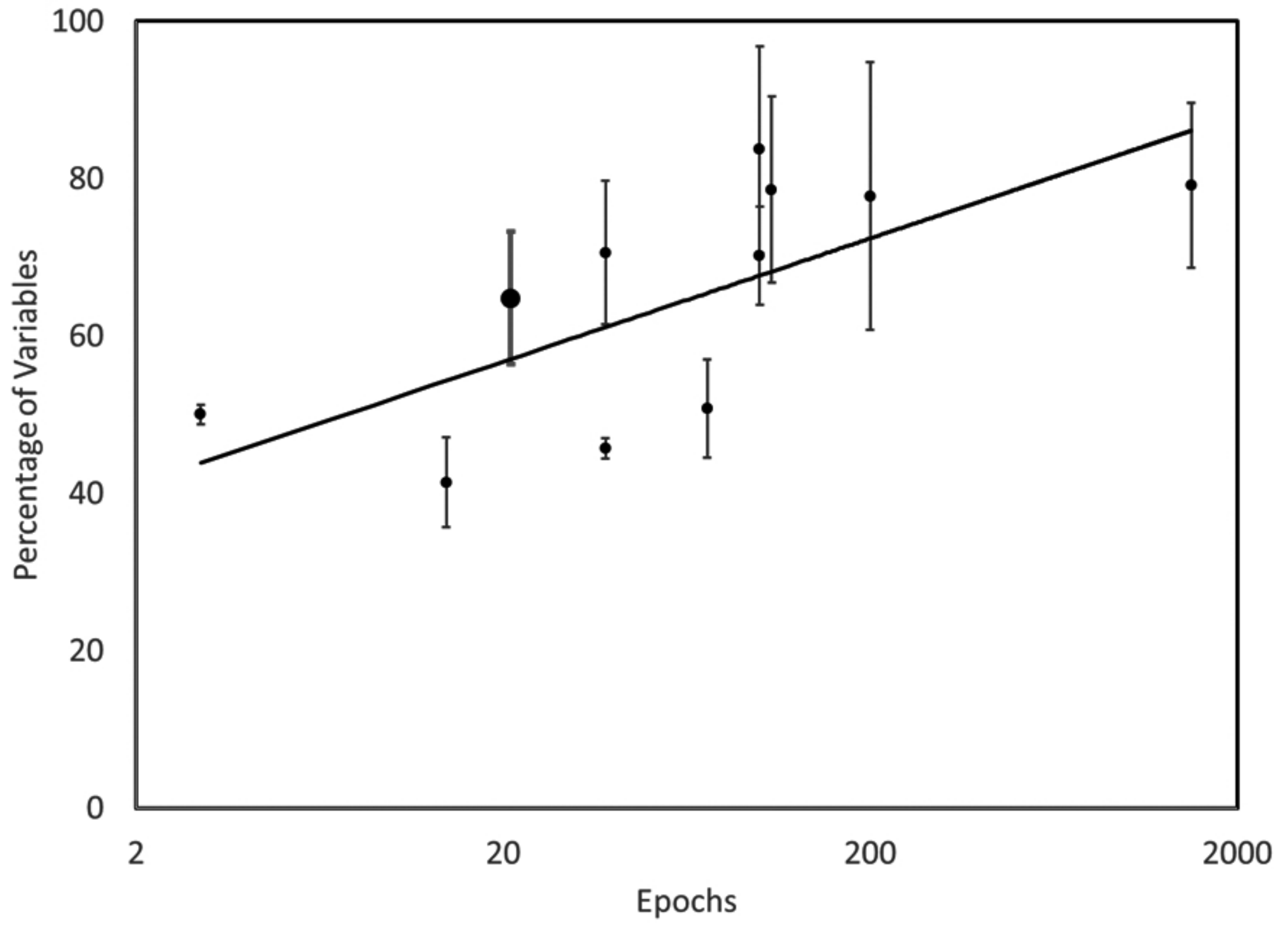}
\caption{Percentage of cluster members identified as variables vs. number of epochs of observation. Data are in order of increasing epochs,  from \citet{megeath2012, deoliveria2008, carpenter2001, wolk2018, rebull2015, wolk2013a, poppenhaeger2015, gunther2014, flaherty2016, parks2014}. The value from this work 
is shown as the large dot with heavy error bars.
\label{epochs}}
\end{figure}

Variability is an alternative indicator of cluster membership that can be applied to very faint detection limits. 
We demonstrated in the preceding subsections that $\sim$ 65\% of the cluster members 
are detected to vary in our survey. This result is placed in context in Figure~\ref{epochs}; roughly, the percentage of variables identified in a cluster increases logarithmically with the number of epochs of observation. Our result is a bit above the overall trend and is consistent with up to $\sim$ 80\% of the cluster members being variable, to be revealed with a sufficiently large number of epochs of observation. Although these statistics give a measure of the incompleteness that will result (i.e., about 35\%), we have an adequate set of observations to identify members down to the brown dwarf regime (late M at the age of Tr 37). Selection of members on this basis should provide an incomplete but unbiased sampling of the cluster membership for sources fainter than our maximum brightness limits of 11.7 in $J$, 11.5 in $H$, and 11.2 in $K$. Given the extinction and relevant range of spectral types, the meaningful brightness limit is 11.2 at $K$; cluster members of this brightness will fall below the limits at $J$ and $H$. The completeness limits at $K$ can be determined from the distribution of variability range in the identified variable sources and the rms noise as a function of $K$ magnitude. The completeness corrections from the identified variables to the total over the range for the brighter sources are negligible for $K < 16$ and are $<$ 20\% for $K < 16.5$.

However, some of the variable stars in the survey may not be members of Tr 37. We will therefore use a color-magnitude diagram (CMD) to distinguish variable stars likely to be members from variable interlopers. The applications of CMDs to other young clusters have established where YSOs should locate in the diagram. In this work, we compare the CMD of Tr 37 with that of IC 348 \citep{luhman2003,luhman2016} for two reasons. First, IC 348 and Tr 37 are thought to be similar in age \citep[$\sim$3 Myr,][]{herbst2008,sicilia-aguilar2005}, so their member stars should fall along similar loci. Second, IC 348 is much closer ($311 \pm 32$ pc \citep{boyce2019} vs. 945$^{+90}_{-73}$ pc \citep{sicilia-aguilar2019}) than Tr 37, with members known down to brown dwarfs. Our UKIRT data on Tr 37 are significantly deeper than the available data on IC 348  and reach a similar range of absolute magnitudes, providing a good comparison.

\begin{figure}
\epsscale{1.2}
\plotone{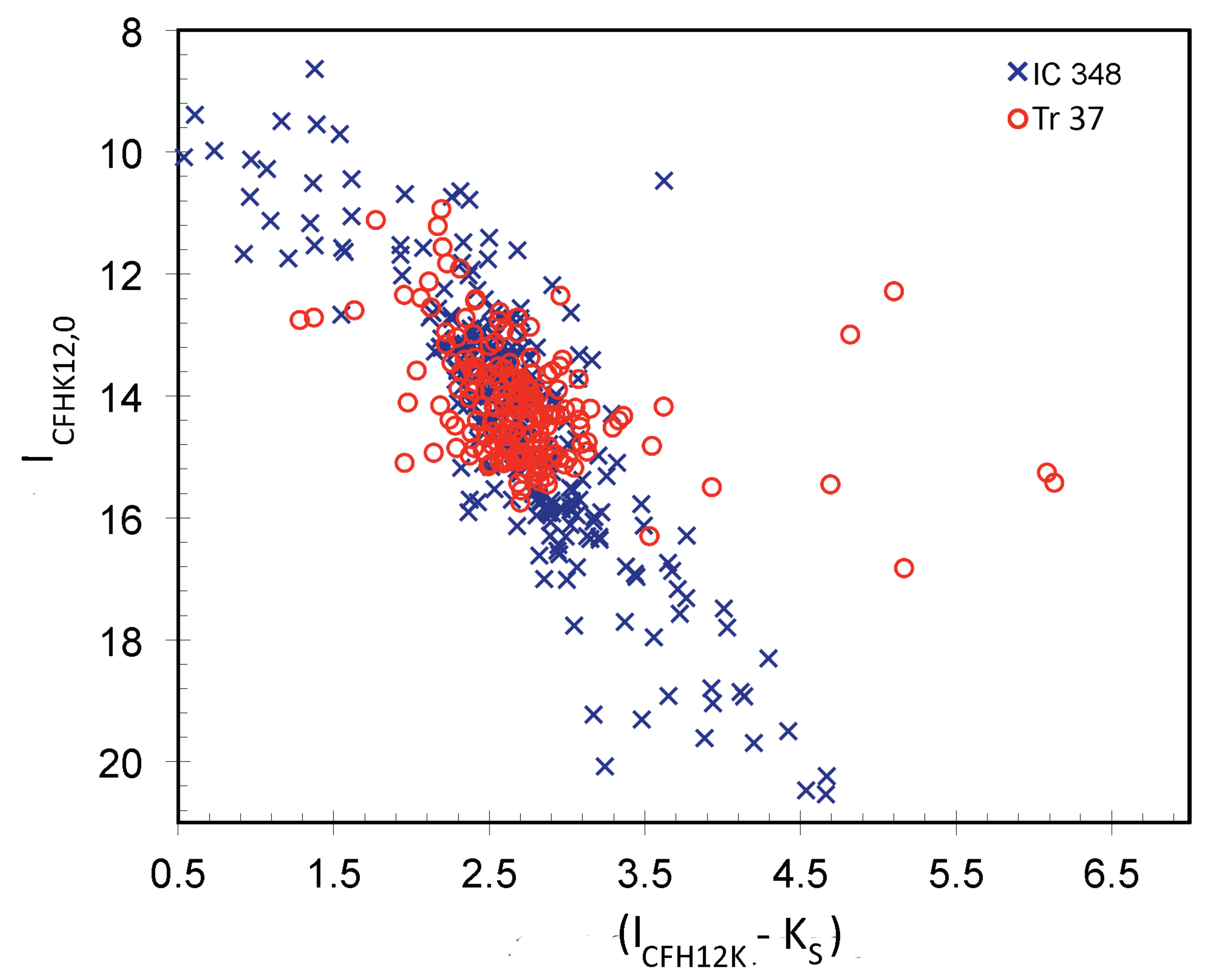}
\caption{$I, I-K$ color-magnitude diagram for all the high probability Tr 37 cluster members from \citet{sicilia-aguilar2013}, superimposed on the CMD for IC 348 from \citet{luhman2016}. The values for IC 348 have been shifted fainter by 2.0620 mag.
\label{IIK-CMD}}
\end{figure}

The CMDs of IC 348 were recently discussed in detail by \citet{luhman2016} based on UKIDSS and CFHT $IZY$ band data and the 2MASS data in $K_S$ band. Since our new UKIRT data are in $JHK$ bands only, for direct comparison we adopt the optical photometry from the Pan-STARRS DR1 catalog \citep{chambers2016}, which, like our photometry on the mosaic images, averages over a number of epochs. For comparison with the IC 348 CMD, we convert the Pan-STARRS $i$ and $z$ magnitudes to $I$ as described in the Appendix. We plot all the spectroscopically confirmed Tr 37 member stars from \citet{sicilia-aguilar2013} together with the IC 348 members from \citet{luhman2016} in the same CMD (Figure~\ref{IIK-CMD}).

\begin{figure}
\epsscale{1.2}
\plotone{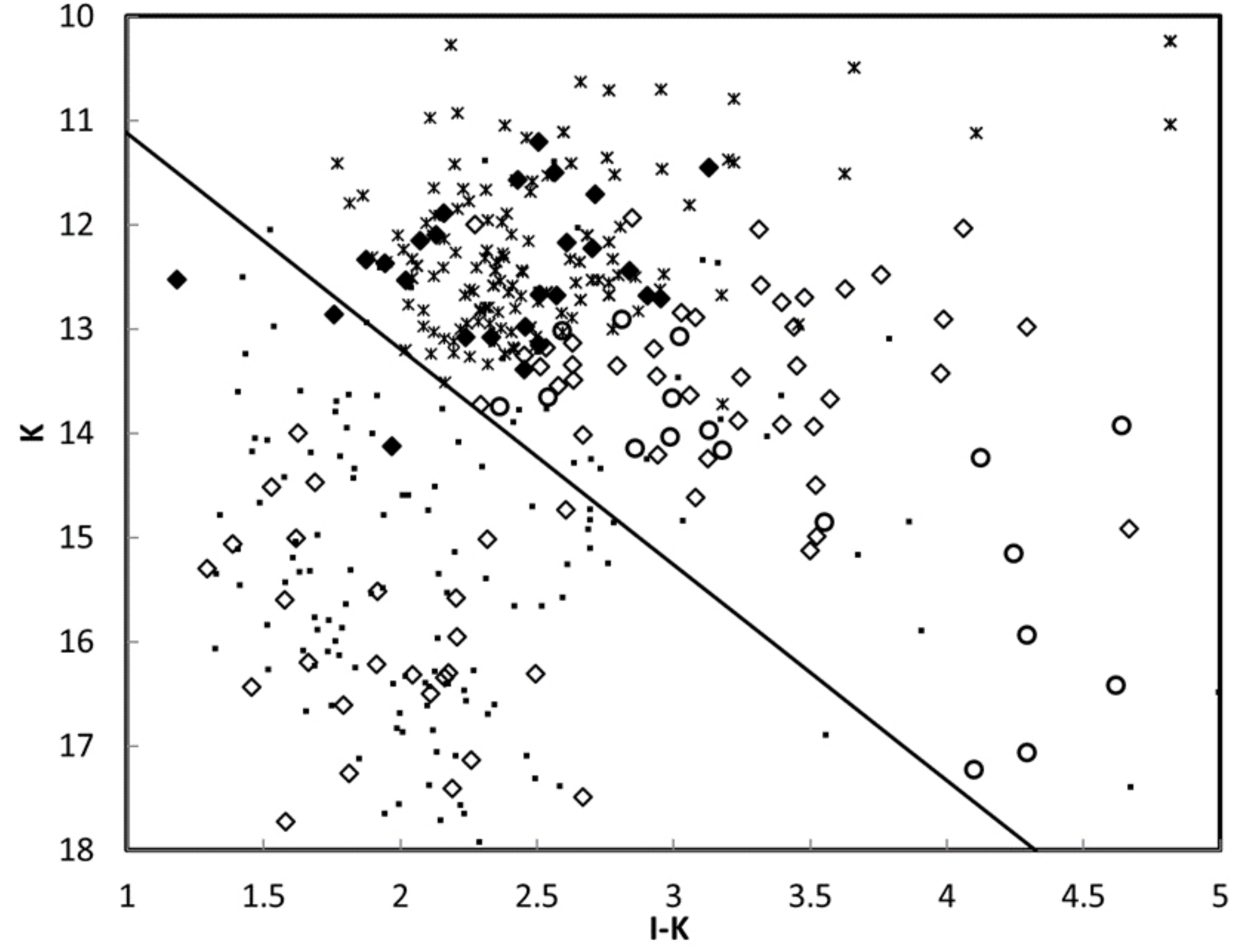}
\caption{$I, I-K$ color-magnitude diagram. The asterisks are for all the high probability Tr 37 cluster members from \citet{sicilia-aguilar2006b}. The filled diamonds are for variable stars discovered in this work and with high weight (SNR $>$ 10) Gaia DR2 parallaxes indicating cluster membership. The open diamonds are similar except that they have low weight Gaia parallaxes (SNR $>$ 2) that are consistent with cluster membership. The small dots are sources where the parallaxes are inconsistent with cluster membership. The open circles are stars without measured parallaxes that we have assigned cluster membership based on their positions on the CMD. The diagonal line is the expected boundary between members and non-members; it has been determined to have the same slope and to be placed similarly relative to the cluster members as in \citet{luhman2016}.
\label{CMD}}
\end{figure}


\begin{figure}
\epsscale{1.2}
\plotone{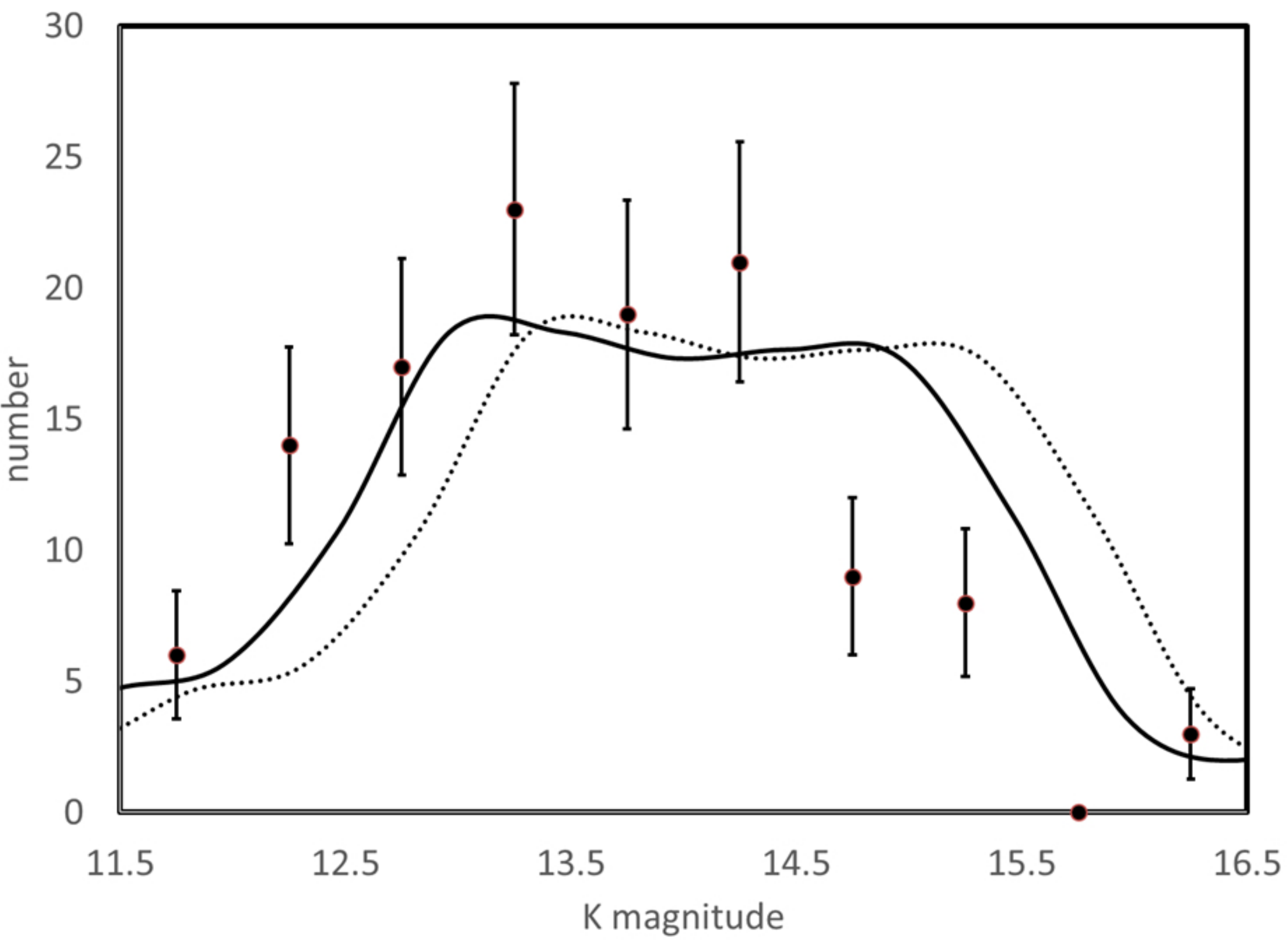}
\caption{The $K$-band luminosity function for IC 348 \citep{muench2003} adjusted to the distance of Tr 37 (dotted line), based on Gaia DR2 parallaxes for both clusters, and compared with the $K$-band distribution of the sample of variable members of Tr37 (points with error bars). The values for Tr37 are not corrected for extinction, which would move them $\sim$ 0.11 - 0.15 magnitudes brighter (this correction has not been made to display the results over the range directly demonstrated not to suffer from biases due to saturation or incompleteness). As shown by the solid line, the fit is significantly improved by shifting the IC 348 luminosity function  0.36 magnitudes brighter, suggesting that the population in Tr37 is somewhat younger than that in IC 348.
\label{K-Lum}}
\end{figure}

We will use the CMD to identify variable stars that could be members of Tr 37. We separated the locus of possible cluster members from that for background stars in analogy with the approach for IC 348 by \citet{luhman2016}. In Figure~\ref{CMD}, we show the CMD for our variable stars with a dividing line placed to be analogous to that used by \citet{luhman2016}. The behavior of the high-probability cluster members from \citet{sicilia-aguilar2006b}  causes us to conclude that most stars above the line are likely to be members. However, we tested all candidate members against Gaia DR2 parallaxes and rejected two as being probable foreground stars. There are 16 stars that meet our criteria for cluster membership but do not have parallax measurements: given the small number of foreground stars among the stars with parallaxes, it is not likely that any of these stars are foreground. We therefore  accept the sources above this line (except the two with large parallaxes) as possible members. The resulting 119 candidate variable cluster members are listed in Table 3, along with their numbers in the membership lists compiled by \citet{errmann2013} and whether they are listed by \citet{sicilia-aguilar2013}.  This sample is likely to be 
about 2/3 complete and, based on the small fraction of field stars that are variable in our data, to  have very low contamination. 

Table 4 lists the additional variable stars that did not pass our tests for cluster membership.

\subsection{K-band luminosity function and CMD Comparison}

Figure~\ref{K-Lum} shows the expected $K$-band luminosity function for Tr 37 based on that of IC 348 \citep{muench2003} adjusted for the relative distance of the two clusters. The similar ages would suggest that the luminosity functions should be similar and indeed they have similar shapes, but the best fit finds that the function is shifted significantly brighter for Tr37. To quantify this effect, we have shifted the luminosity function to minimize $\chi^2$, which occurs at a shift of 0.36 magnitudes, or at $\sim$ 0.5 magnitudes if we make a correction for extinction. 

This conclusion can be tested by comparing the two clusters on the CMD. If the two clusters have similar ages, their members would be expected to locate around the same locus. We make a linear fit for the locus of IC 348 and use the same fixed slope to force a fit to the members of Tr 37. The difference in the intercept, $2.06 \pm 0.36$, is the {\it relative} distance modulus from isochrone fitting. However, the Gaia parallaxes indicate that the true distance modulus is about 0.36 mag greater than this value, agreeing qualitatively with the shift  estimated from the luminosity function (before the additional correction for extinction). 

The shift in the $K$ luminosity function has the greater weight over the CMD because of its smaller dependence on extinction; nonetheless, the shift becomes about 0.5 magnitudes with the extinction correction for Tr37. The nominal errors in the Gaia-estimated distance \citep{sicilia-aguilar2019} are too small to explain these shifts. However, there is evidence for a small bias toward underestimating parallaxes in Gaia \citep[see summary in] []{xu2019}. A weighted average of all the measurements \citep{xu2019} yields $-53 \pm 3$ $\mu$as, but this value is strongly influenced by the single input with the smallest quoted error. The average omitting this value is $-56 \pm 11$ $\mu$as. If we adopt $-55 ~\mu$as, the relative distance modulus between IC 348 and Tr37 is reduced by 0.11 mag with a 95\% confidence error of about $\pm 0.3$ mag, i.e., to 
account for the 0.5 mag shift in the K luminosity function would require that the distance be overestimated by an amount that is outside the expected errors.  The behavior suggests that the K- and M-stars in our sample, which dominate the luminosity function, are somewhat younger than similar stars in IC 348. Our variability-selected sample is centered around IC 1396A (see Figure 4), so a plausible explanation is that many of these stars are associated with a population that is younger than typical for Tr37, e.g., the very young stars ($\sim$ 1 Myr old) discussed by \citet{getman2012}.

\subsection{Characteristics of Individual Sources}

\begin{figure}
\epsscale{1.2}
\plotone{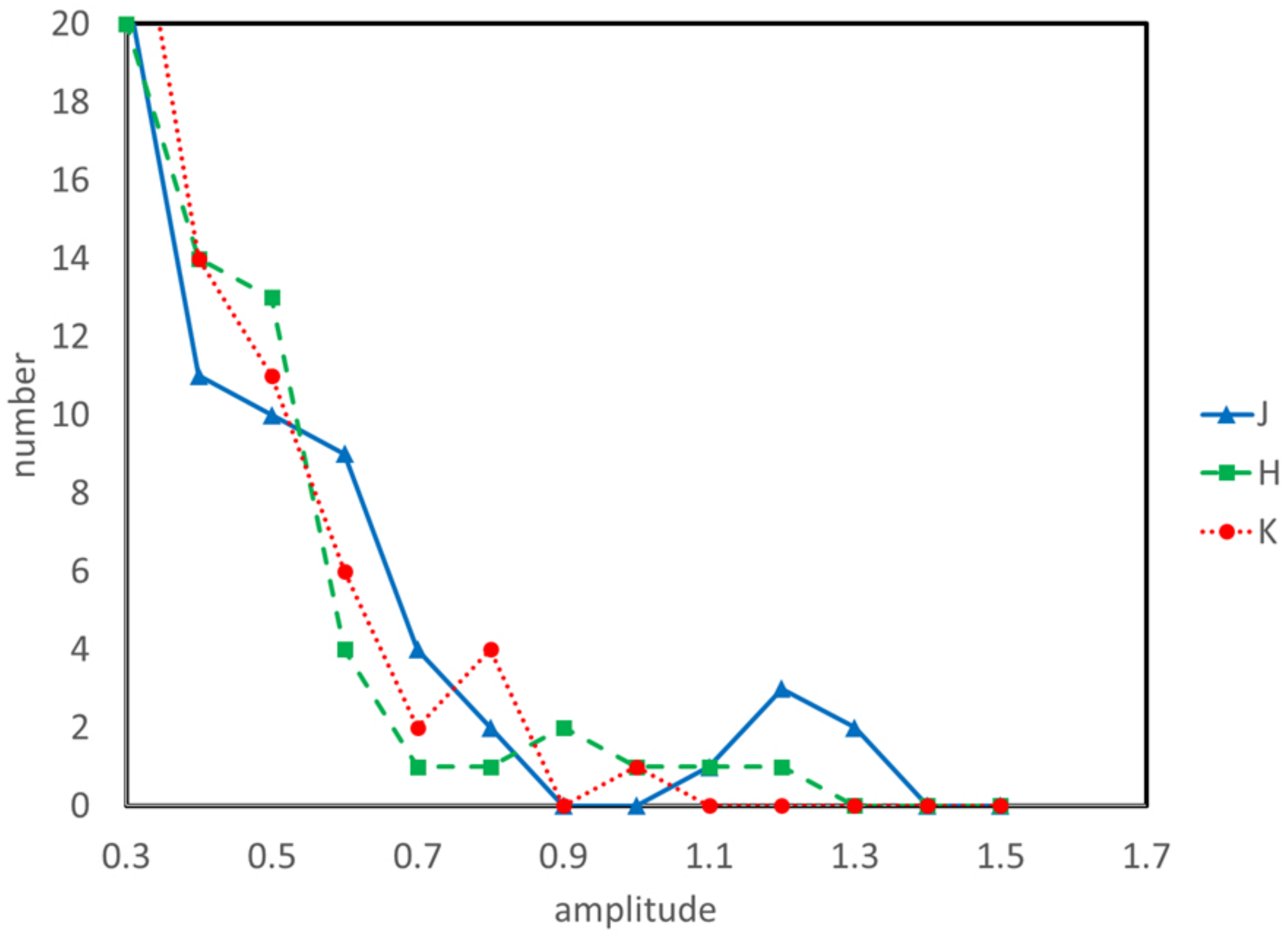}
\caption{Peak-to-peak variability of the 119 probable cluster members for $J$, $H$, and $K$ bands.  There are more highly variable sources (amplitude $>$ 0.9 mag) at $J$ and $H$ than at $K$. 
\label{Var_amp}}
\end{figure}

This section illustrates the variability patterns of individual sources. By necessity, we focus on those with relatively large amplitude variations so that the patterns emerge clearly.  This criterion will tend to omit stars that vary because of cool spots but should include the other previously identified types of variables \citep{wolk2013a, wolk2013b}. We should detect many periodic variables \citep{rice2012, rice2015}, although the low cadence of our observations will keep us from identifying them as such.


\subsubsection{Accretion hot spots on the stellar surface}

\begin{figure}
\epsscale{1.2}
\plotone{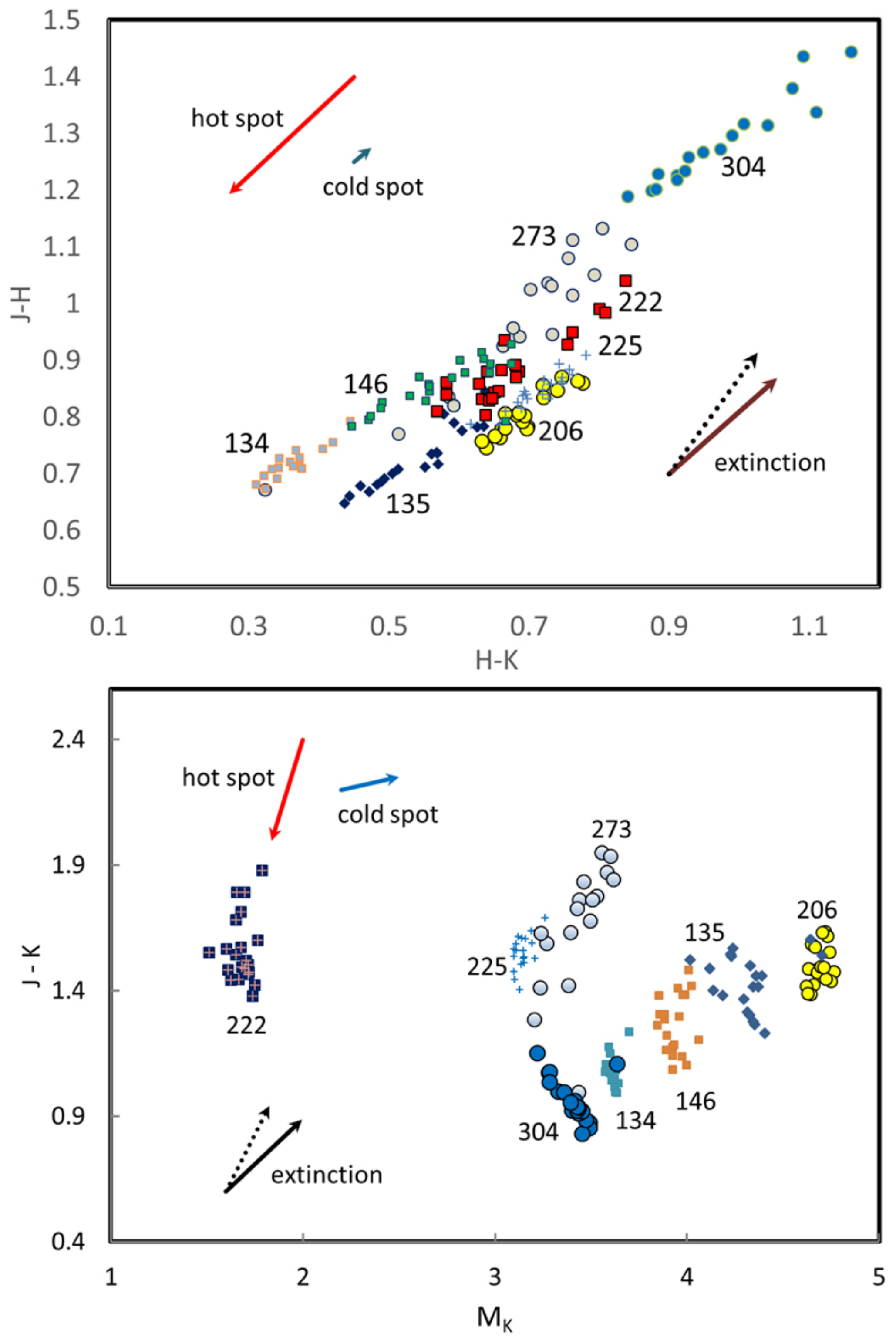}
\caption{The $J-H$, $H-K$ color-color diagram (CCD) and $J - K$, $M_K$ color magnitude diagram (CMD) for sources where the change in $J$ is large, but there is relatively little change at $K$. The behavior of a source acquiring a hot or cold spot is shown at the upper left with a red or blue arrow, respectively. The behavior due to changes in extinction is shown in the lower right of the CCD and lower left of the CMD, with a dotted line for standard interstellar behavior ($A_V = 2$) and a solid line for an extinction law as expected for large grains (shown to be characteristic of the variable extinction in these sources in Section 5.6.3). The behavior of these sources suggests the presence of accretion hot spots, which are well above the temperature of the stellar photosphere and thus produce a large change at the shorter wavelengths when they rotate into view.   
\label{fig12e}}
\end{figure}

\begin{figure}
\epsscale{1.2}
\plotone{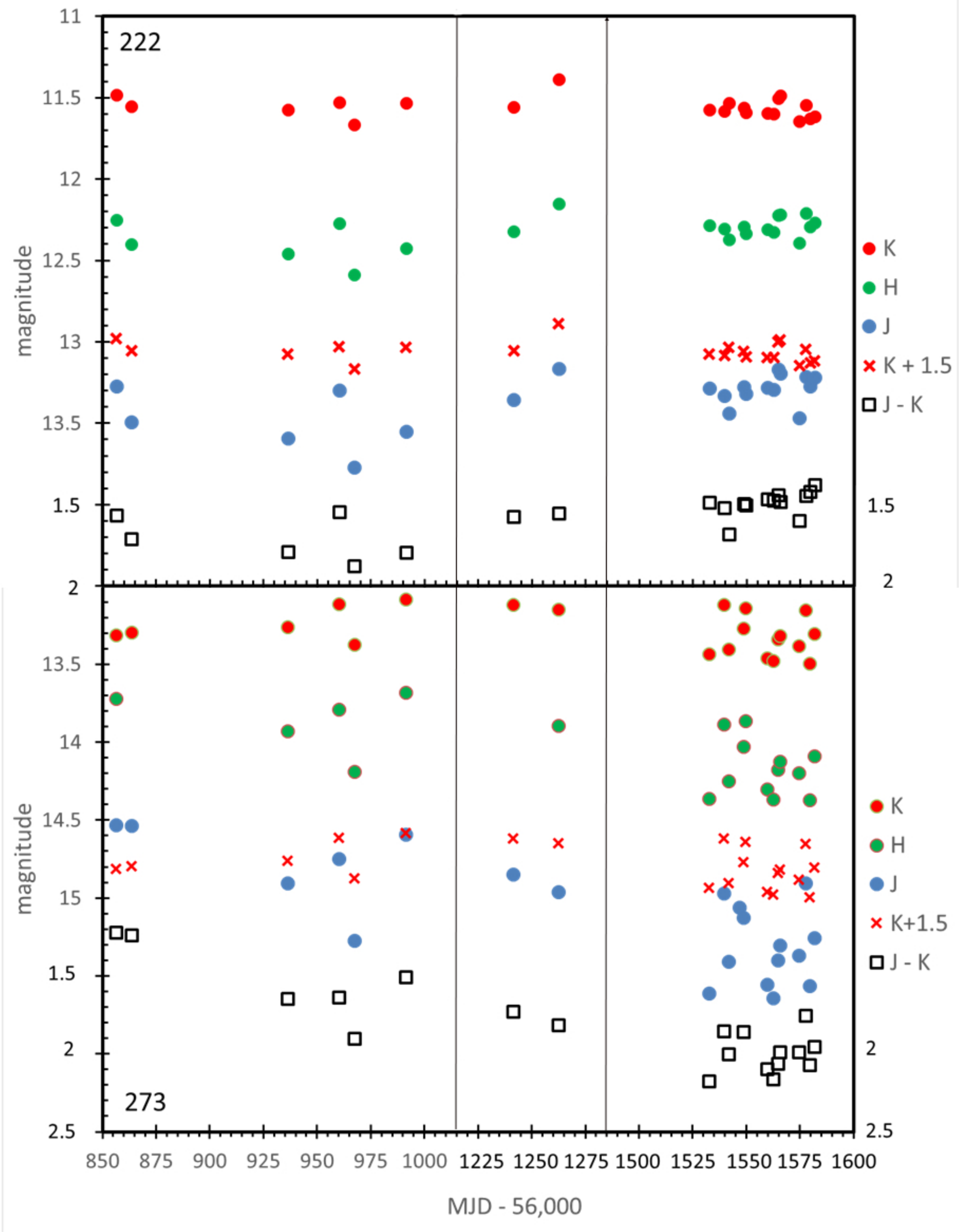}
\caption{The light curves for sources 222 and 273, two of the most variable sources where the amplitude at $J$ is $>$ twice as large as that at $K$. The light curves have been compressed by eliminating epochs with no data, indicated by the light vertical lines. The data at $K$ are repeated 1.5 magnitudes fainter to facilitate comparison with the $J$ lightcurve. Comparison of the shifted $K$ curve and the $J$ one, and also the $J - K$ color, show that the sources become bluer when they are brighter, i.e., the variations are driven by relatively large increases in the output at $J$. 
\label{fighotspts}}
\end{figure}

\begin{figure}
\epsscale{1.2}
\plotone{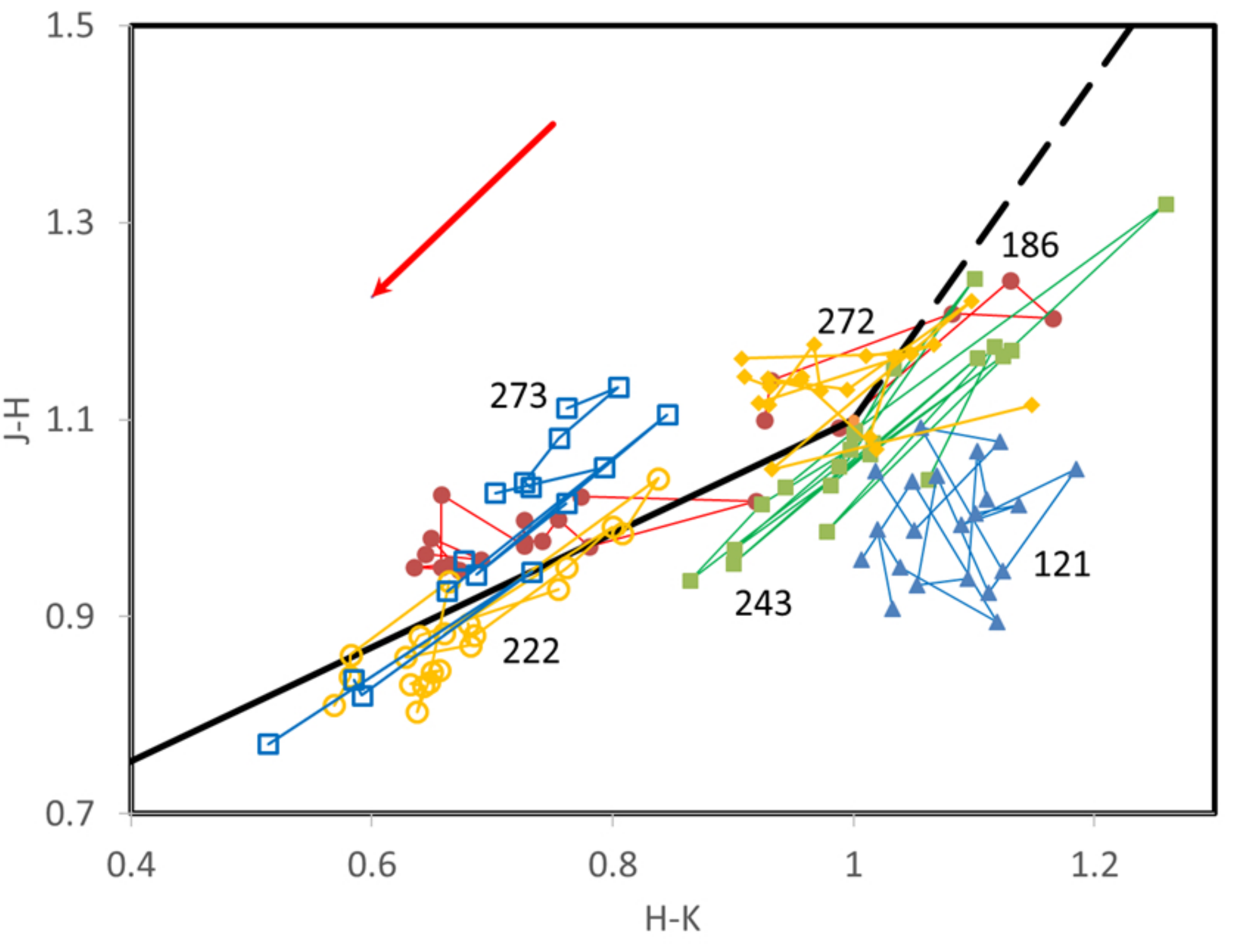}
\caption{The behavior of the two hot-spot prototypes numbers 222 and 273, and four EXor-like sources, 121, 186, 243, and 272 on the $J-H$, $H-K$ diagram. The red arrow at the upper left is the expected vector for a hot spot and the black line (solid and dashed) is the CTTS locus from \citet{meyer1997}. Sources 222 and 273 follow the CTTS locus with moderate scatter, supporting the hypothesis that their variations arise primarily due to episodic accretion events onto the stellar surfaces, i.e., hot spots.  The EXor candidates have a more chaotic behavior and more extreme colors. 
\label{figccdacc}}
\end{figure}

This section discusses the sources whose behavior indicates accretion hot spots, based both on the color behavior of the variations, their chaotic nature, and the placement if the sources along the classical T Tau locus in the $J-H$, $H-K$ color-color diagram.  Figure~\ref{Var_amp} shows that more cluster members have a high degree of variability in $J$ than do in $H$ or $K$. As shown in Figure \ref{fig12e}, these objects also tend to have substantial scatter in the $J-H$, $H-K$ diagram. Light curves for two examples can be found in Figure~\ref{fighotspts}. These objects potentially change output because of the appearance of a hot spot due to localized accretion onto the stellar surface \citep[e.g.,][]{wolk2013b}. To provide a specific example, we have used the high flux level for source 222 on MJD = 56,960.192 and the low one a week later, on MJD 56,967.20.  We have dereddened the $JHK$ measurements for $A_V$ = 1.3, the average value for the cluster \citep{errmann2013}, converted the resulting magnitudes to flux densities, and then subtracted the values for MJD 56,960.19 (low) from those for MJD 56,967.20 (high) to provide three photometric points for the difference, which we can ascribe to the variable component. Any fits are very degenerate; for example, we find that the photometry in the low state can be fitted well by a blackbody of 1950K, and to fit the high state we need to assume a coverage of 1.2\%\ of its surface by a hot spot at 5200K. This calculation is used to generate the hot spot vector in Figure~\ref{fig12e}. If we increase the extinction by a factor of 3 to $A_V$ = 3.9, the temperatures become 2500K and 9000K, respectively. The latter two values are reasonably representative of the photospheric temperature of low mass pre-main-sequence stars with hot spots observed in other studies, e.g., \citet{wolk2013b}\footnote{In the following section we will identify some sources with dramatic rapid variations that change relatively little in near infrared color.}. The rapidity of some of the variations (e.g., source 273) may result from hot spots rotating in or out of sight. Four of these sources, numbers 174, 222, 273, and 304, have been studied previously and the latter three have been found to have strong H$\alpha$ emission \citep{nakano2012, sicilia-aguilar2013}, consistent with the suggestion that their variability is associated with accretion, e.g., hot spots. Fig~\ref{figccdacc} shows sources 222 and 273 on the $H-K$, $J-H$ color-color diagram, along with the classical T Tauri Star (CTTS) locus \citep{meyer1997}. The behavior of lying close to this locus throughout their variations indicates that these two sources are experiencing high and variable mass accretion and have similar accretion disk geometries (i.e., inner hole radii) to CTTS's in general \citep{meyer1997}.

To complement the hot spot vector in Figure~\ref{fig12e}, we show the vector for normal interstellar reddening \citep{rieke1985} and for reddening by large grains \citep{steenman1989}. Extinction will be discussed further in Section 5.6.3. We also show the expected behavior for cold spots \citep{wolk2013b}. These vectors show that in the $J-H$, $H-K$ diagram, the behavior of hot spot variations may be difficult to distinguish from extinction changes, but the $J-K$, $M_K$ diagram is more diagnostic, particularly relative to extinction by large grains, which we show in Section 5.6.3 is common around young stars.

\begin{figure}
\epsscale{1.2}
\plotone{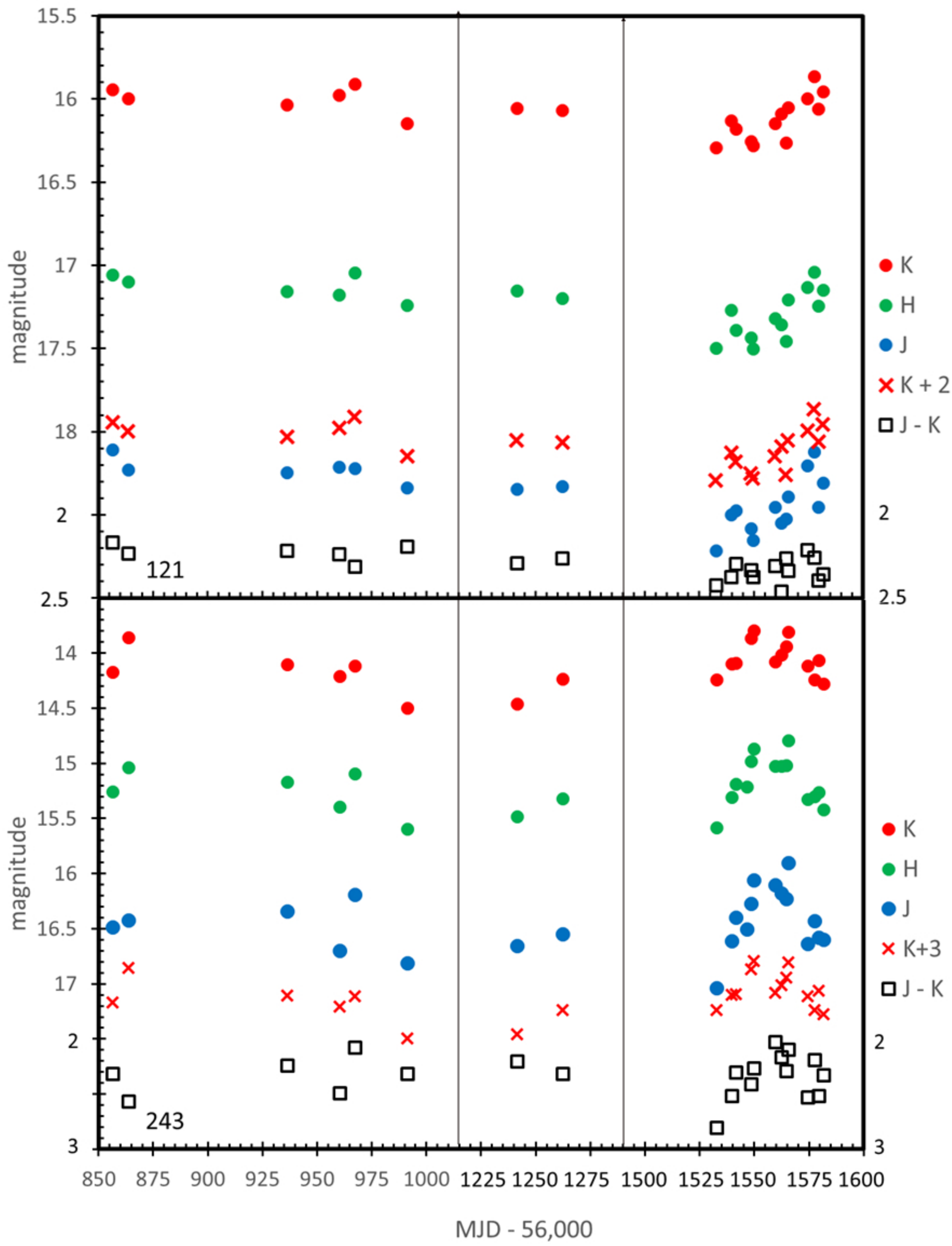}
\caption{The light curves for sources 121 and 243 compressed as in Figure~\ref{fighotspts}. Source 121 has fluctuations that are relatively similar in amplitude at all the bands, whereas source 243 has variations driven by increases in the output at $J$ as indicated by it becoming bluer when it is bright.
\label{figexorsa}}
\end{figure}

\begin{figure}
\epsscale{1.2}
\plotone{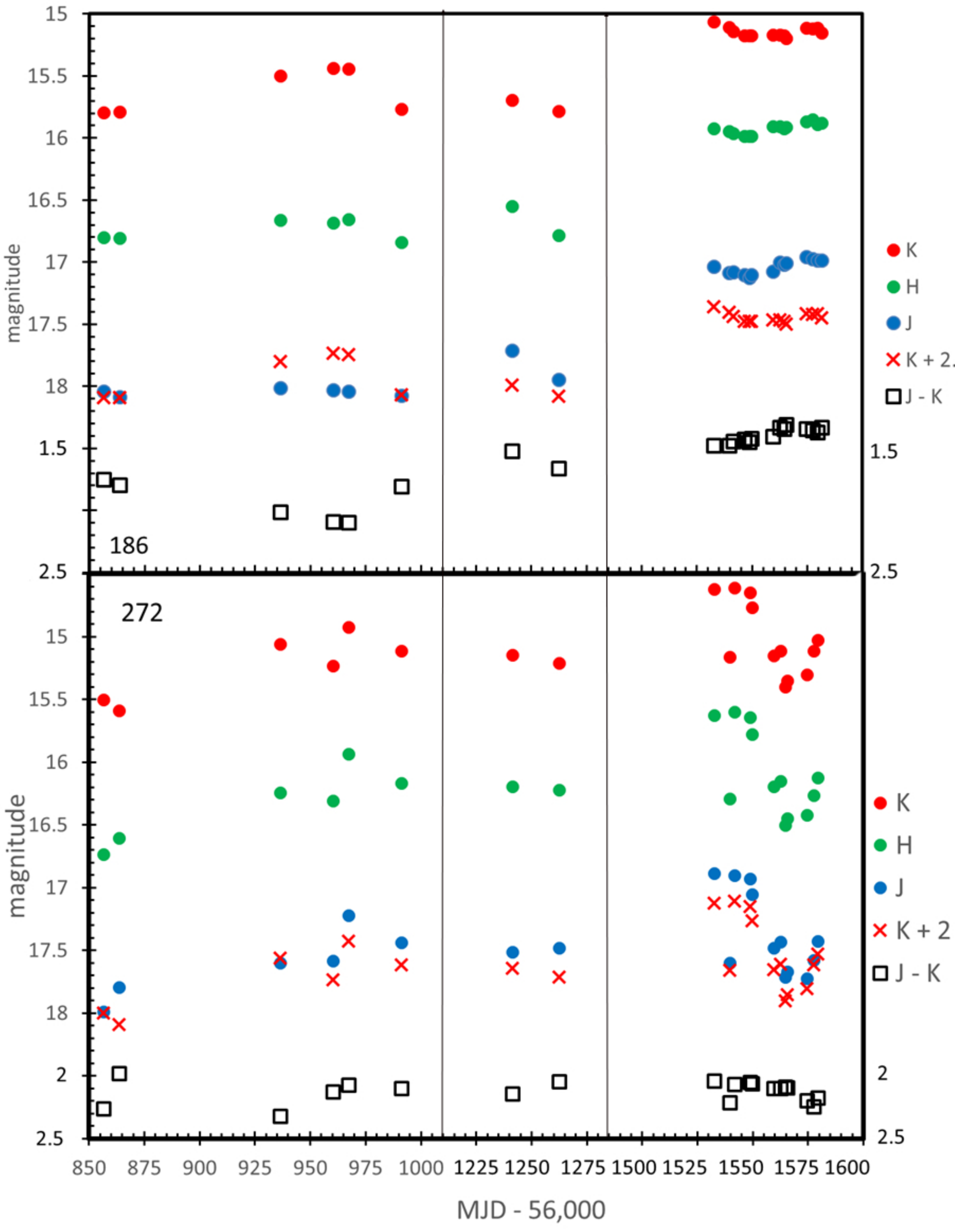}
\caption{The light curves for sources 186 and 272 compressed as in Figure~\ref{fighotspts}. Source 186 became significantly bluer after its transition to a relatively steady bright state. The colors of Source 272 are relatively unchanged through its variations.  
\label{figexorsb}}
\end{figure}

\begin{figure}
\epsscale{1.2}
\plotone{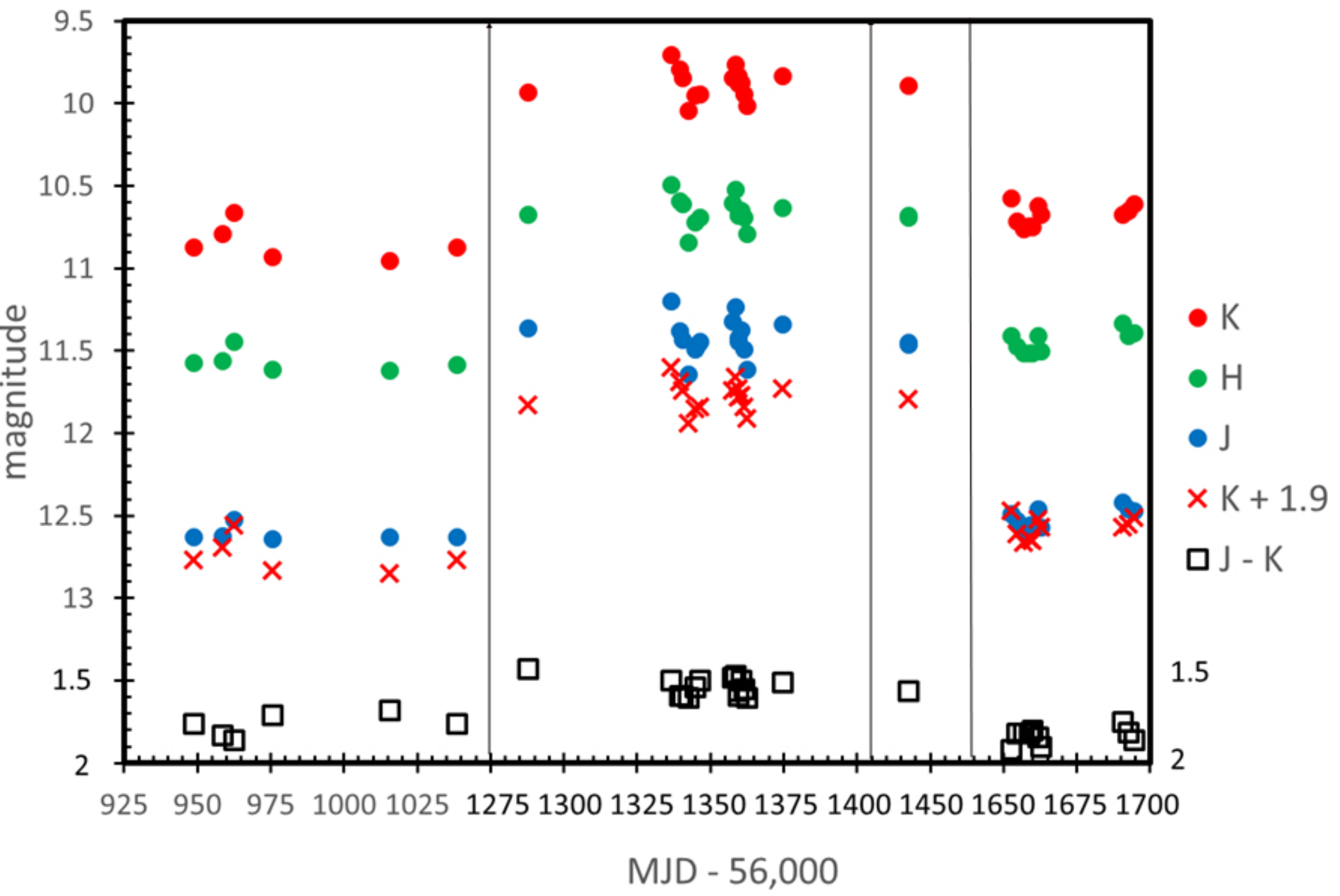}
\caption{The light curve of V1118 Ori through its outburst in 2015 - 2016, compressed as in Figure~\ref{fighotspts}, from \citet{giannini2016, giannini2017}. Its dramatic outburst results in a near infrared color only slightly bluer than when it is at a lower state, a situation similar to some of the EXor-like sources in Tr 37.
\label{v111ori}}
\end{figure}

\subsubsection{Chaotic variability and EXors}

Some of the most extreme variability patterns and amplitudes resemble those of EXor type objects (named after EX Lupi); four examples are sources 121, 186, 243, and 272 (Figure~\ref{figexorsa} and Figure ~\ref{figexorsb}).  
The amplitudes of these objects are similar to those of, e.g., sources 222 and 273, but their changes seem to be subject to more chaotic behavior and longer term trends. 
EXors show eruptive variability caused by rapid accretion rate changes by orders of magnitude (see Herbig 1998 and Audard et al. 2014 for review), and show smaller amplitude variability with more frequent outburst 
episodes compared with the related FU Ori type objects; the latter show much larger variability ($\Delta~V \gtrsim 4$~mag) that can last for several decades.
Frequent variability episodes of EXors are likely related to magnetospheric accretion of circumstellar disk materials onto 
the young star, which show higher amplitude variability at shorter wavelengths.  
\citet{giannini2016, giannini2017} show multi-wavelength variability for an EXor, V1118 Ori;  
we reproduce their data in Figure~\ref{v111ori} showing the 2015 - 2016 outburst of this star, 
for comparison with the variations of Sources 121, 186, 243, and 272.  
The variability patterns of V1118 Ori and our sources 121, 186, and 243 are similar with a higher $J$ band variability 
amplitude than in $H$ and $K$ band; for source 272, the changes are more similar in amplitude across the three bands. Peak to peak variability amplitudes of source 121 are 0.61, 0.46, and 0.43, for source 186 they are 1.13 mag, 0.99 mag, and 0.73 mag, for source 243 they are 1.14 mag, 0.8 mag, and 0.7 mag, and for source 272 they are 1.10 mag, 1.13 mag and 0.98 mag in $J$, $H$, and $K$ band respectively. In all four cases, the variations have trends with month- or year-long time scales, consistent with EXor behavior \citep{audard2014}. 
 Source 186 shows in Figure~\ref{figexorsb} a different pattern than sources 121, 243, and 272, namely a general brightening with modest variations otherwise.  
Although this behavior is still consistent with being an EXor, it could also arise through removal of 
obscuring materials along the line of sight (after MJD 57,500).

These four sources are shown on the $H-K$, $J-H$ diagram in Figure~\ref{figccdacc}. Their extreme nature is confirmed by their passing into the protostellar disk regime to the right of the CTTS locus. 

Another set of sources show chaotic variations but with smaller amplitudes (0.3 - 0.4 mag); we term these chaotic variables. Examples are numbers 72, 76, 157, 166, 232, 251, and 252. Their behavior might, for example, result from minor accretion events, or from a combination of effects, such as both hot spots and variations of extinction (see following subsection). In such cases our simple categorization would not be definitive. 

\subsubsection{Large grains in circumstellar disks}

\begin{figure}
\epsscale{1.2}
\plotone{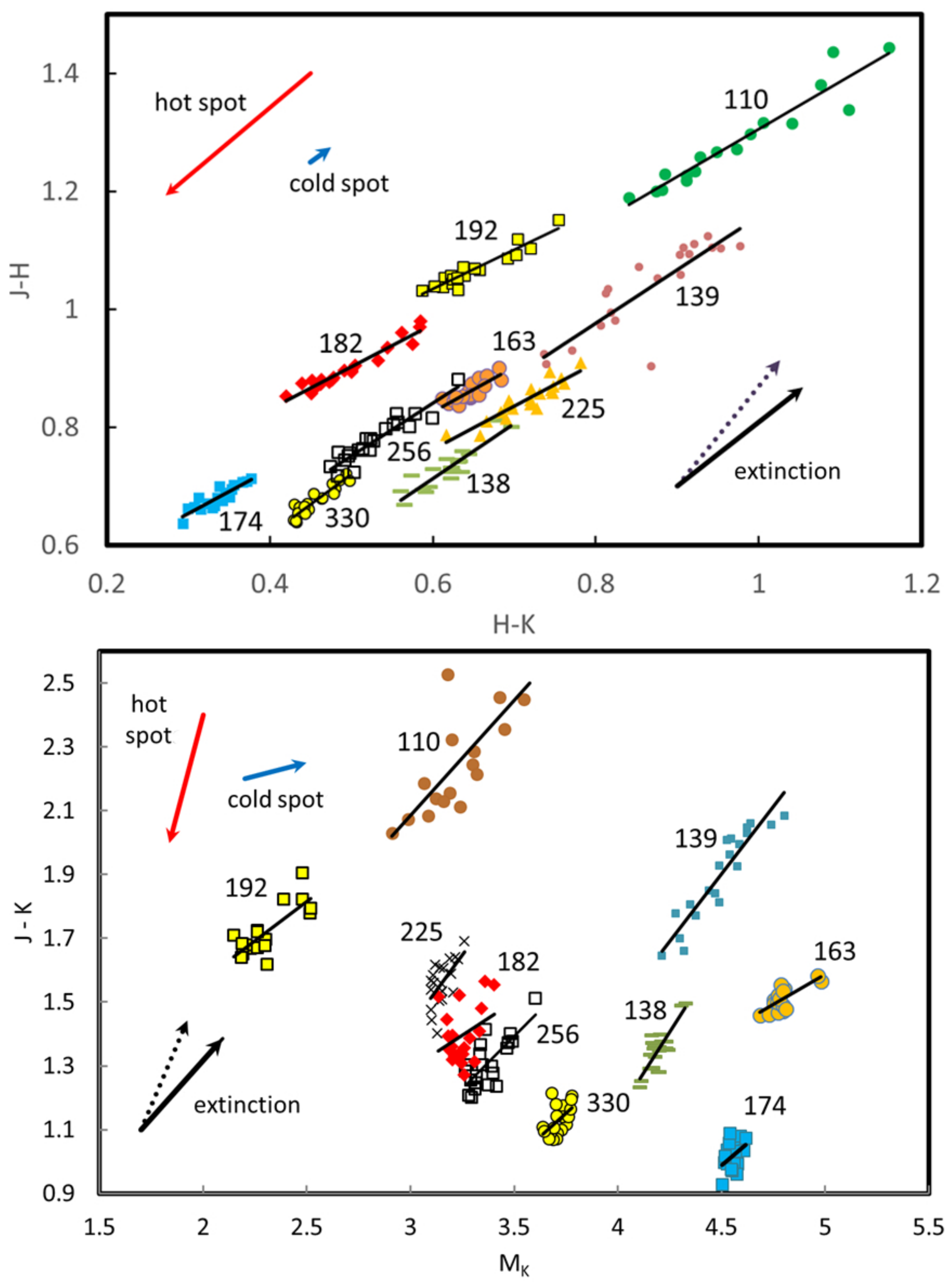}
\caption{This diagram shows eleven sources where the pattern of change is consistent with variations in the amount of extinction, e.g. from inhomogeneities in a circumstellar disk. The standard interstellar extinction behavior \citep{rieke1985} for $A_V = 2$ is shown as a dotted black line, and the extinction for a  particle size distribution extending upward to $\sim$ 1 $\mu$m \citep{steenman1989} is shown as a solid black line. The behavior of a source acquiring a hot or cold spot is shown at the upper left with a red or blue arrow, respectively. The thin black lines show fits by linear regression to the data for each source.
\label{figlrggrns}}
\end{figure}

Some sources appear to vary due to changes in extinction caused by grains larger than those typical of interstellar dust. Figure~\ref{figlrggrns} shows color-magnitude and color-color diagrams for eleven sources with this behavior. We suggest that the variability in these sources is driven primarily by the intervention of clumps or other structures in their circumstellar disks \citep{rice2015, stauffer2015}. The large-grain vectors are based on the work of \citet{steenman1989}, who showed that a simple extinction model with lower and upper size limits on the grains of 0.005 and 0.22 $\mu$m, respectively, and with a size distribution of 
$n(a) \propto a^{-3.5}$, gives a good fit to the normal extinction law and then explored the changes resulting from an increase in the upper size limit. The solid black vectors shown in Figure~\ref{figlrggrns} are from their calculation for a ratio of total to selective extinction, R = 5.9, corresponding to an upper size limit of 
$\sim$ 1 $\mu$m.

\begin{figure}
\epsscale{1.2}
\plotone{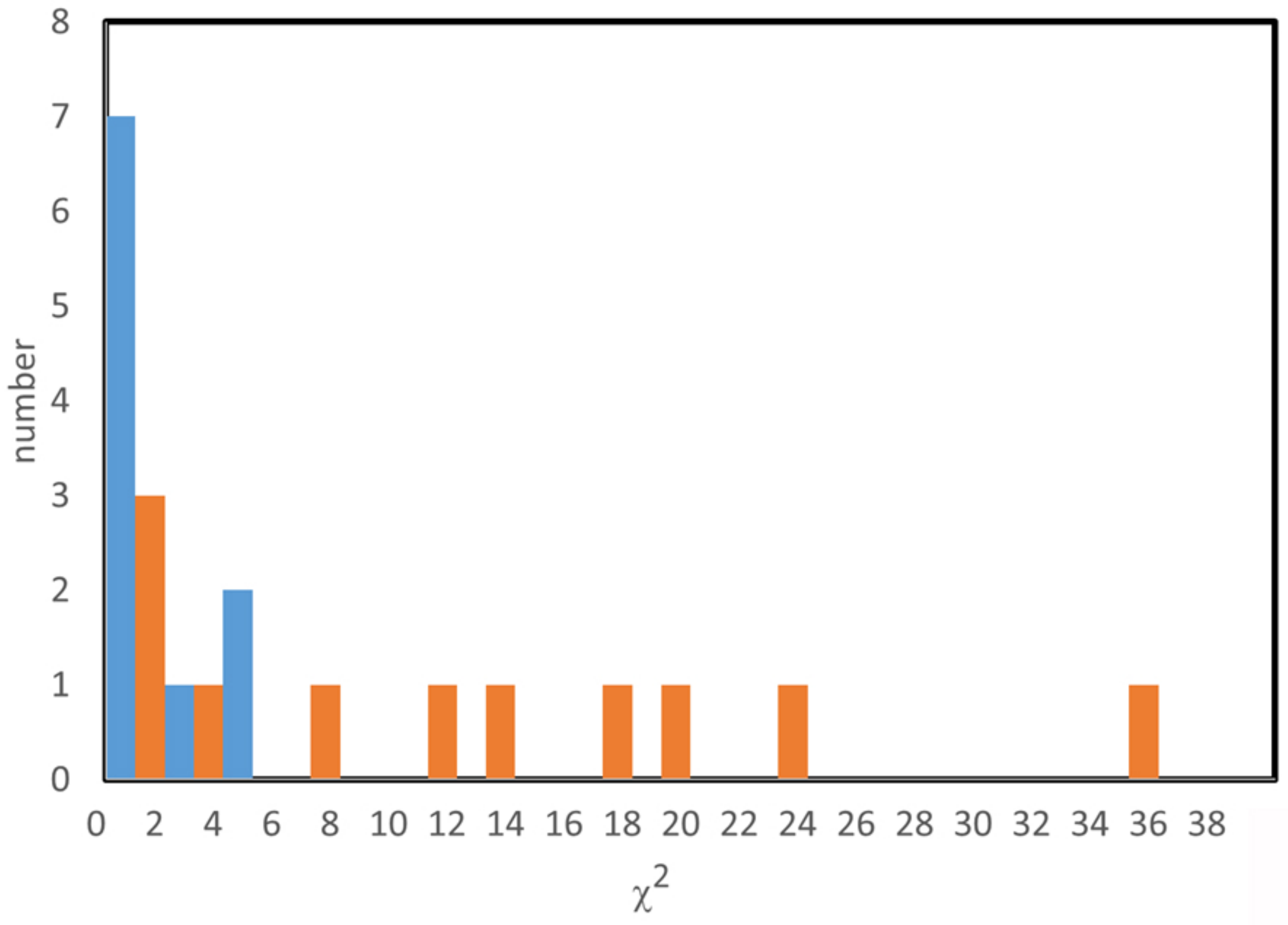}
\caption{To separate sources where the variations may be dominated by extinction and those with variations dominated by accretion, we compute the reduced  $\chi^2$ relative to the relation $(H-K) = 0.77 (J-H)$ + constant. The blue bars are the sources used to study the extinction behavior (Figure ~\ref{figlrggrns}) and the orange bars are those we suggest are dominated by accretion events (Figures \ref{fig12e} and \ref{figccdacc}).
\label{figchisq}}
\end{figure}

We expect that extinction will result in a well-determined linear relation between $J-H$ and $H-K$ with minimal scatter around this trend as the extinction varies. We use this expectation to separate the sources into ones that obey this expectation and those that have larger scatter. To identify candidates where the variations might be dominated by variable extinction, we set the slope in ($H-K$) vs. ($J-H$) equal to the median for the sources in  this part of our study, 0.77, and then minimized the reduced $\chi^2$ for a linear fit to the data, with results as shown in Figure ~\ref{figchisq}. We examined the sources in the intermediate zone, with values of $\sim 2 - 4$ individually, and rejected those with anomalous behavior in the $M_K$ vs. $(J-K)$ diagram, typically ones with small slopes with large scatter, and those with relatively small variations. The sources with behavior possibly dominated by extinction are shown in blue in Figure ~\ref{figchisq} and the remaining ones in orange.  Although our procedures to distinguish hot spot variations from extinction ones might just have separated sources with different slopes in the $(H-K)$ vs. $(J-H)$ diagram,  Figure ~\ref{figchisq} demonstrates that, instead, it separates sources with small scatter around the linear trend from those with large scatter.

\setcounter{table}{6}
\begin{deluxetable*}{lcc}
\tabletypesize{\scriptsize}
\renewcommand\thetable{9}
\tablecaption{Slopes by linear regression from Figure~\ref{figlrggrns} compared with those for other extinction determinations}
\tablewidth{0pt}    
\tablehead{
\colhead {Source} &
\colhead {E(J-K)/$\Delta$ K}  &
\colhead {E(H-K)/E(J-H)} 
}
\startdata

\citet{rieke1985}, R = 3.1  & 1.50 $\pm$ 0.2 & 1.70 $\pm$ 0.15 \\
\citet{harris1978}, $\rho$ Oph R $\sim$ 5  &   & 1.40 $\pm$ 0.15  \\
\citet{cardelli1989}  & 1.48  & 1.22  \\
\citet{he1995}, general  &  &  1.64 $\pm$  0.26   \\
\citet{he1995}, $\rho$ Oph  &  &  1.64 $\pm$ 0.23  \\
\citet{steenman1989}, R = 5.9  &0.72  &  1.09  \\
\hline
69  & 0.72 $\pm$ 0.15  & 0.81 $\pm$ 0.07 \\
90  & 1.01 $\pm$ 0.17  & 0.93 $\pm$ 0.10 \\
109  &  0.37 $\pm$ 0.09 & 0.77 $\pm$ 0.13 \\
116  & 0.57 $\pm$ 0.31 & 0.76 $\pm$ 0.10 \\
124  & 0.43 $\pm$ 0.27 & 0.72 $\pm$ 0.05 \\
126  & 0.50 $\pm$ 0.09 & 0.67 $\pm$ 0.05 \\
145  & 0.90  $\pm$ 0.27 & 0.74  $\pm$ 0.07 \\
168  & 0.66  $\pm$ 0.14 & 0.90  $\pm$ 0.08 \\
207  & 0.59  $\pm$ 0.21 & 1.00  $\pm$ 0.10 \\
median &  0.59   &   0.77  
\enddata

\end{deluxetable*}

Changes in the minimum grain size have little effect on the relative extinction in the $JHK$ bands \citep{steenman1991}, in agreement with expectations since the minimum size of typical interstellar grains is far smaller than the wavelengths of these bands. This situation underlies the independence of the $JHK$ extinction law on the selective extinction ratio, $R_V = E(B-V)/A_V$ \citep{cardelli1989}. The behavior of the sources in Figure~\ref{figlrggrns} and as summarized in Table 7 therefore suggests that the  dust grains responsible for the extinction have a larger maximum size than those in the general interstellar medium, i.e. in agreement with other indications that in these protoplanetary disks significant grain growth has occurred.  

\subsubsection{Possible eclipses}

\begin{center}
\begin{figure}
\epsscale{1.1}
\plotone{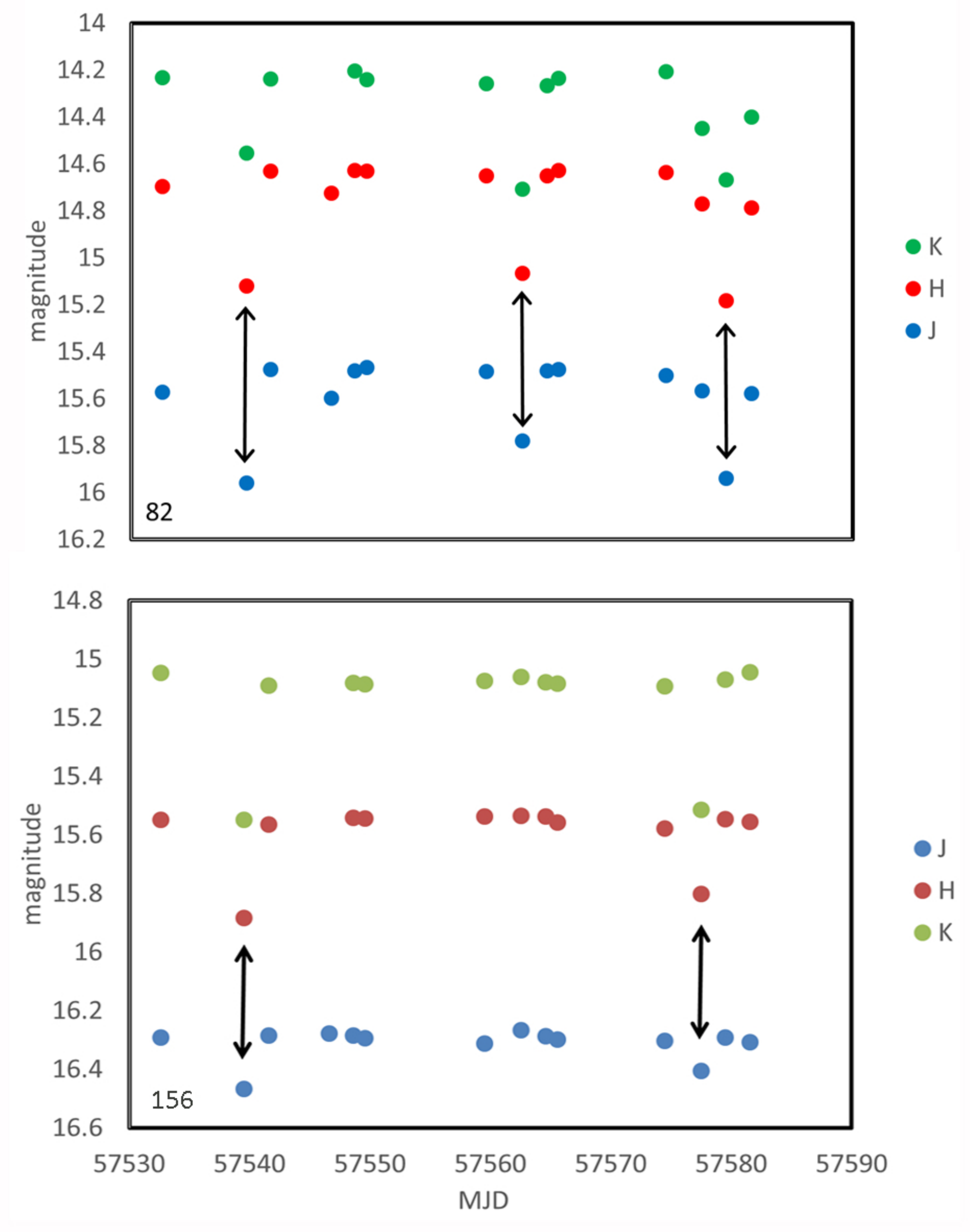}
\caption{Examples of light curves suggestive of eclipses. The $K_S$ photometry is in green, 
$H$ in red, and $J$ in blue, and the arrows indicate possible eclipses.
\label{eclipsesa}}
\end{figure}
\end{center}

Figure~\ref{eclipsesa} shows the relevant parts of the light curves of two sources where eclipses are a possibility. Other possibilities for such events are Sources 95, 137, 146, 288, 304, and 347.  These eight sources should be observed with a more rapid cadence to test this possibility.

These events could either be a result of the intervention of a luminous object, i.e., a companion star, or that these sources are ``dippers" \citep{morales-calderon2011}, behavior characterized by \citet{rice2015} for sources in Orion and attributed by them to the intervention of a compact dust cloud. We can distinguish the two possibilities on a general basis: dippers would usually have a stronger event the shorter the wavelength, whereas stellar eclipses would have a similar number of events with stronger or weaker events with shorter wavelength (depending on whether the hotter or the cooler member of the pair was diminished). Of the eight candidate objects, five have events at equal strength in all three bands, two (137 and 156) have stronger events the longer the wavelength, by about a factor of two from $J$ to $K$, and one has a modestly stronger event at $J$ than at $K$ (a factor of 1.5). This distribution favors the hypothesis that they are mostly due to stellar binaries, although sources 137 and 156 could also be dippers.

\subsubsection{Intrinsic K-band variability and instabilities in circumstellar disk rims}

\begin{center}
\begin{figure}
\epsscale{1.2}
\plotone{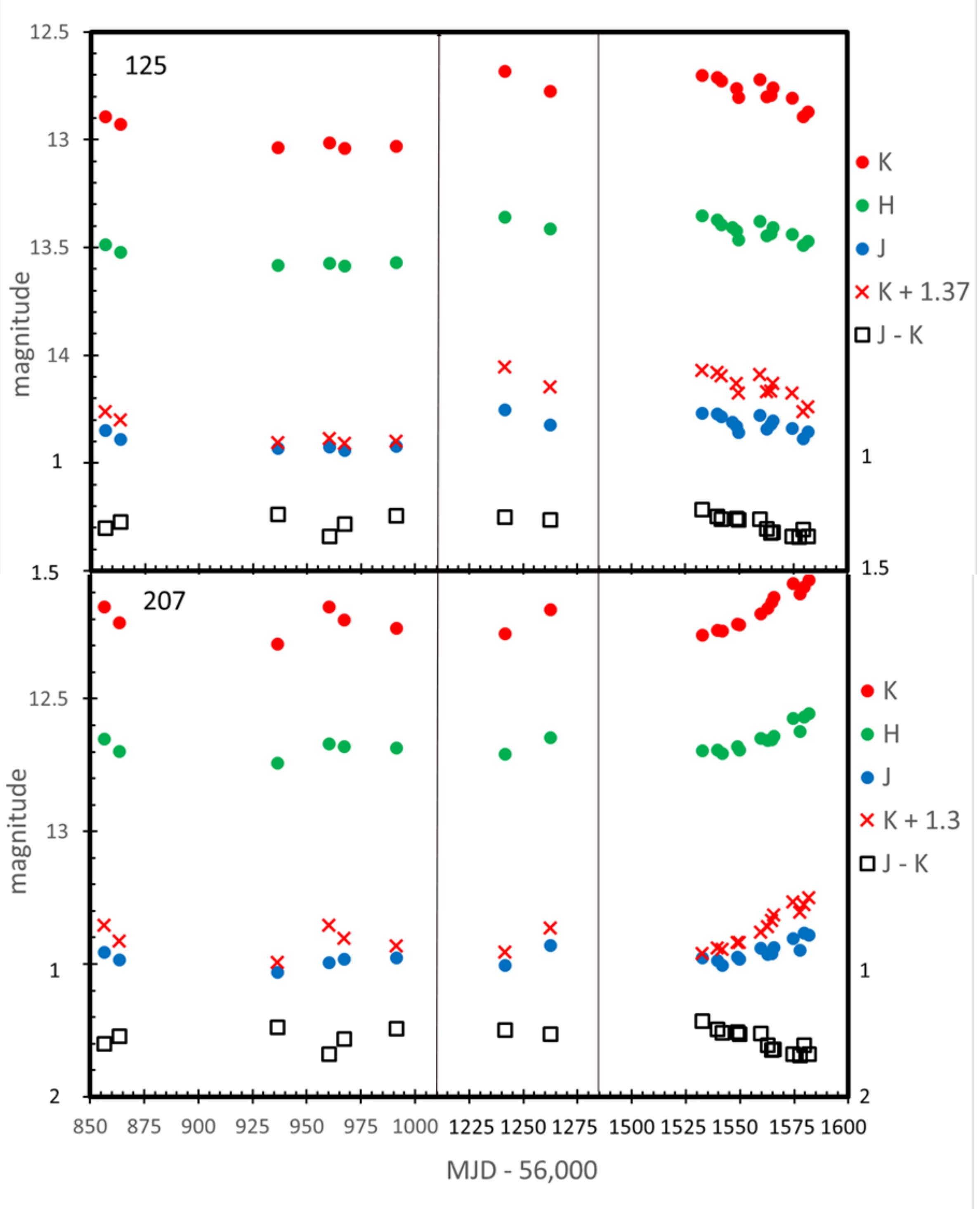}
\caption{Example sources with large intrinsic variability at $K$, reducing toward shorter wavelengths. The $K$ photometry is repeated fainter as indicated to facilitate comparison with that at $J$.
\label{figkband}}
\end{figure}
\end{center}

\begin{center}
\begin{figure}
\epsscale{1.2}
\plotone{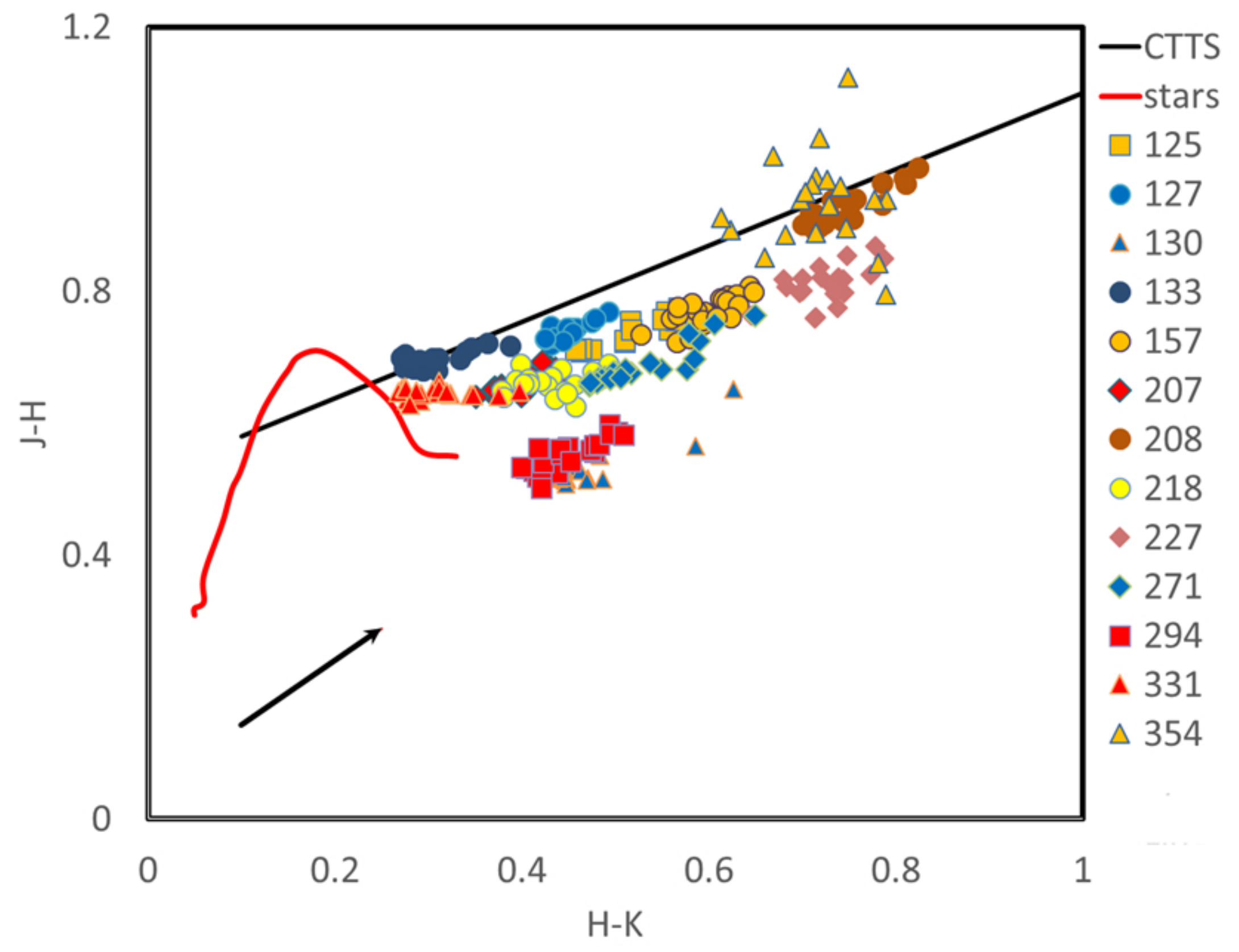}
\caption{Placement of predominantly $K$ variables on the $H-K$, $J-H$ diagram. The stellar color locus (red line) is from \citet{luhman2010}, supplemented by standard colors for types earlier than K4; the black solid line indicates the CTTS locus from \citet{meyer1997} and the black vector shows the effect of extinction by large grains.  
\label{figkccd}}
\end{figure}
\end{center}

Another class of source has the larger variations in the $K$ band, with decreasing amplitudes for $H$ and then $J$, which we will show is indicative of grains being exposed to direct stellar irradiation after being lifted out of an optically thick circumstellar disk. Two examples are shown in Figure~\ref{figkband}. The placement of these sources on the $H-K$, $J-H$ diagram is shown in Figure~\ref{figkccd}; they tend to fall below the CTTS locus but parallel to it and in a position that is consistent with their having greater excesses in $K$ above the photospheric emission (to the right of the stellar locus) and/or reduced hot-spot type activity (down relative to the CTTS locus)\footnote{It is not uncommon for stars to be distributed around the CTTS locus rather than all being above it \citep[e.g.,][]{hillenbrand1998}.}. 

Mid-infrared variations in young stars can arise purely from instabilities in the circumstellar disk \citep[e.g.,][]{muzerolle2009, flaherty2016}. We will test whether such instabilities can account for the variability in our sample where the $K$-band changes are larger than those at $J$ and $H$. The near infrared emission should originate from dust at the inner disk rim, a position regulated by the grain sublimation temperature, as is confirmed by near infrared  interferometry and reverberation mapping \citep[e.g.,][]{millan-gabet2007, dullemond2010, anthonioz2015, meng2016}. The $JHK$ colors of CTTS also often show an excess attributed to emission by the inner disk \citep[e.g.,][]{cieza2005}. The sublimation temperature is conventionally placed at 1500K, appropriate for small silicate grains. However, the process is complex, depending on (1) grain composition; (2) ambient gas density; and (3) size and time of exposure. Specifically, the carbonaceous grain components sublimate at $\sim$ 2000K. In addition, when silicates break down some of the products, e.g., FeO and MgO, have similarly high sublimation temperatures \citep{mann2007}. As shown in detail by \citet{baskin2018}, the sublimination process is one of erosion and hence the sublimation temperature is increased for grains immersed in relatively dense gas.  As they also show, it takes longer to eliminate large grains, allowing transient events involving large grains (see their Section 4.6.1) to exhibit higher temperatures than for steady-state conditions with small ones. 

\citet{cieza2005, mcclure2013} show that the near infrared excesses of CTTSs can be fitted with blackbodies of temperature $1750 \pm 250$K. For comparison, we have fitted blackbodies to the {\it variable} emission of the dominant $K$-band-variability sources, as listed in Table 8. To obtain measurements of the variable component of the emission we subtracted the measurements in a low state from those in a high one. Specifically, we took the differences in brightness on the indicated dates (averages over a range of dates if so indicated in the table) and corrected for the extinction of Tr 37 ($A_V$ = 1.3, \citet{rieke1985} extinction law). The fits were done by $\chi^2$ minimization assuming a net error in the differences of 5\% (including the non-statistical contribution estimated by \citet{rice2015}) and the indicated errors in the temperatures are $\sim \pm$ 200K. It is the nature of the fits to have the same offsets relative to the $J$ and $K$ measurements, they can be considered to give the $J$ to $K$ color temperature with an additional check that the $H$ point is consistent.  With the exception only of Source 331, satisfactory fits were achieved, that is deviations of $\lesssim$ 10\% from the fit for the individual bands, up to $<$ 20\% in two cases. All but one case indicates a temperature in the range appropriate for transiently heated dust from the circumstellar disk. The  exception (Source 130) is for the only source where we differenced data over an interval of about a year; the other cases are all for intervals of order a month, i.e., for transient events with that timescale. 

\begin{deluxetable*}{llccccc}
\tabletypesize{\footnotesize}
\renewcommand\thetable{10}
\tablecaption{Temperatures fitted to variable components}
\tablewidth{0pt}    
\tablehead{
\colhead {Source} &
\colhead {MJD\tablenotemark{a}}  &
\colhead {J (mag)} &
\colhead {H (mag)} &
\colhead {K (mag)} &
\colhead {T (K) A$_V$ = 1.3} &
\colhead {T (K) A$_V$ = 2.3} 
}
\startdata
125  &  57533 - 57540  &  14.27  & 13.36  & 12.71  &  --  & -- \\ 
 --  &  57580 - 57582  &  14.40  &  13.48  &  12.88  &  1970  & 2110  \\
127  &  57550 - 57562  &  13.53  & 12.66  &  12.13  &  --  &  --\\
--  &  57574 - 57579  &  13.48  &  12.58  &  12.01  &  1840  & 1950  \\
130  &  57241 - 57262  &  16.10  &  15.35  &  14.66  &  --  & --\\
--  &  57532 - 57582  &  16.12  &  15.45  &  14.93  &  1150\tablenotemark{b} & 1200\tablenotemark{b}  \\
133  &  56960  &  14.00  &  13.14  &  12.67  &  --  & --\\
--  &  56991 & 14.03  &  13.19  &  12.81  &  1590  &  1680 \\
157  &  57533 - 57540  &  14.54  &  13.67  &  13.01  &  -- &  -- \\
--    &  57565 - 57566  &  14.42  &  13.50  &  12.80  &  1850  &  1970  \\
207  &  57533 - 57542  &  13.49  &  12.70  &  12.25  &  --  &  -- \\
--  &  57580 - 57582  &  13.39  &  12.56  &  12.06  &  2130 &  2300  \\
208  &  57242  &  14.40  &  13.29  &  12.40  &  --  & --\\
--  &  57262  &  14.52  &  13.46  &  12.67  &  1570  & 1660 \\
218  &  56960 - 56967  &  12.69  &  11.87  &  11.39  &  -- &  -- \\
--  &  56991  &  12.57  &  11.74  &  11.16  &  2060 & 2210  \\
227  &  56856  &  14.73  &  13.72  & 12.86 &  -- &  -- \\
--  &  56863  &  14.40  &  13.435  &  12.58  &  2410  & 2620  \\
271  &  56936  &  13.80  &  12.96  &  12.29  &  --  &  -- \\
--  &  56967  &  13.57  &  12.70  &  12.03  &  2400 &  2620 \\
294  &  57533 - 57542  &  16.00  &  15.30  &  14.77  &  -- & --  \\
--  &  57566  &  15.945  &  15.22  &  14.63  &  1840  &  1960 \\
331  & 57533 - 57540  &  14.54  & 13.75  &  13.39  &  -- & --  \\
--  & 57562 - 57565  &  14.45  &  13.67  &  13.20  &  1890\tablenotemark{c} & 1890\tablenotemark{c}  \\
354  & 57549 - 57562  &  18.14  &  17.05  &  16.24  &  --  & -- \\
--  & 57566 - 57582 &  18.26  &  17.21  &  16.47  &  1750 & 1860  \\
\enddata
\tablenotetext{a}  {Each source has two entries; the first one is the date or range of dates for the high state, and the second is the date or range of dates
for the low state. The temperatures are fitted to the differences of the fluxes at each band in the high minus the low state.}
\tablenotetext{b} { Time interval is $\sim$ one year compared with $\sim$ one month for the other sources.}
\tablenotetext{c} {  Poor temperature fit because indicated change at $H$ is relatively low.}
\end{deluxetable*}

With the exception of Source 331 (where the fit of a single blackbody was unsatisfactory), the success of these fits and the small range of temperatures indicates that the $K$-band variations are {\it not} correlated with accretion events that create stellar hot spots, since as shown in Section 5.6.1, these events would force the fits to significantly warmer temperatures. Only sources 227 and 271, with fitted temperatures of $\sim$ 2400K, (plus source 331) have any  suggestion of such accretion events. Although we have dereddened the photometry for the average in Tr 37 of A$_V$ = 1.3 \citep{errmann2013}, there is a range of reddening and some estimates are somewhat higher, e.g., from \citet{sicilia-aguilar2013} $1.56 \pm 0.55$. To bound the possibilities, we also computed temperatures after dereddening for A$_V$ = 2.3. The conclusions are only slightly modified: now sources 227 and 271 are lifted out of the range where emission by sublimating dust is plausible. However, it is unlikely that source 271 is so strongly reddened, since that would make its intrinsic $J-H$ color too blue compared with the other similar cluster members \citep{sicilia-aguilar2010} assuming their reddening is average for Tr 37. That is, at least 10 and probably 11 of 13 stars with dominant $K$-band variations show colors consistent with the variable signal arising from dust heated transiently to its sublimation temperature.  

This result is similar to that of \citet{flaherty2014}, who found no correlation between X-ray and mid-infrared variability of pre-main-sequence stars, i.e., that the mid-IR variations arise from the disk and are not related to simultaneous or nearly simultaneous accretion events.  Instead, the behavior indicates a substantial level of turbulence at the inner disk rim, sufficient to expose otherwise protected dust to the radiation of the star. Such behavior is predicted by  a number of theoretical simulations \citep[e.g.,][]{bans2012, turner2014, flock2017}.

\subsubsection{Large variations of neutral color: occultation by the circumstellar disk}

\begin{center}
\begin{figure}
\epsscale{1.2}
\plotone{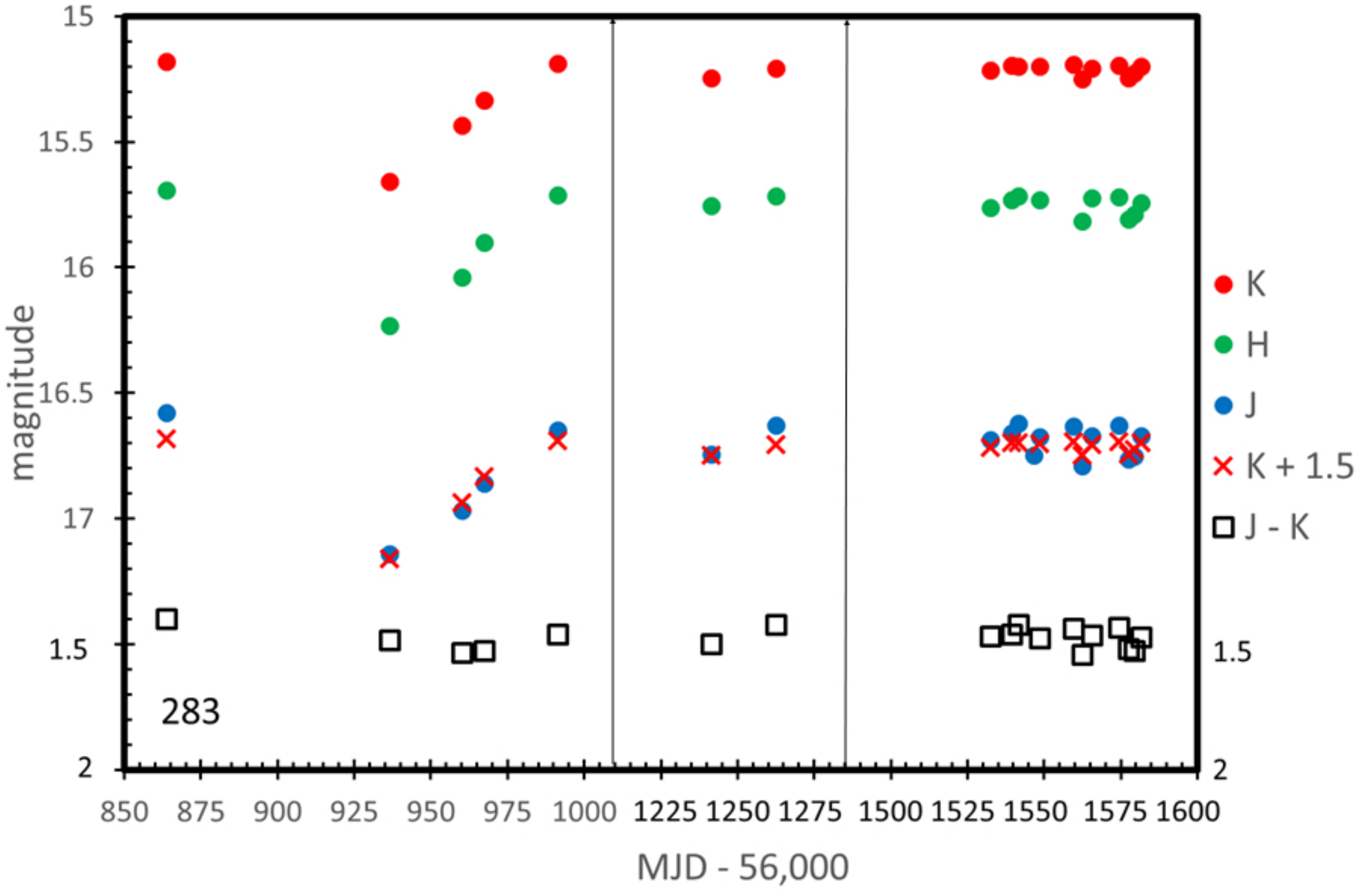}
\caption{Source 283, with a large drop in all colors, otherwise with only modest variations. As shown both by the comparison of the $K$ with the $J$ lightcurve and the $J - K$ color, the changes in this source are neutral in color. This object falls on the CTTS locus at a modest level to the right of the colors of normal young stars. 
\label{fig183}}
\end{figure}
\end{center}

Surveys similar to ours but with higher cadence have shown that there are many sources that vary roughly equally in $J$, $H$, and $K$, with periods of a few days to a few weeks and amplitudes of a few tenths of a magnitude. Because of the aliasing of such behavior into our lower cadence, such sources are likely to appear to have chaotic variations and we do find a significant number of such objects. Since our data are inadequate to study them in any detail, here we call attention to just one more object, source 283,  with very large changes, roughly neutral in color as shown in Figure ~\ref{fig183}. The fitted temperature of the change between the low state on MJD 56,936 to the high state in the last year of monitoring is 2600K, too warm to be thermal emission by the disk but cooler than expected for ongoing accretion.  The excess resembles the broad and featureless continua of a sample of T Tau stars studied by \citet{fischer2011}, which they found could be fitted by blackbody spectra with temperatures between 2200 and 5000K; but see also \citet{mcclure2013} who did not find this component to be necessarily present. 

The $J-H$ color of source 283 averages 0.93 and remains within the errors at this value throughout its change in flux. This color is similar to the average for faint members of Tr 37 from  \citet{sicilia-aguilar2010}, $0.87 \pm 0.10$. The net change of 0.45 mag would require a change in the radiating area by a factor of $\sim$ 1.5, outside the usual hot spot paradigm. Instead, the behavior resembles that of ``faders'' in which a warp or disruption in the circumstellar disk at a significant distance from the star creates a long-duration occultation \citep[e.g.,][]{hamilton2001, bouvier2013}.  One example around GI Tau has a similar duration and recovery behavior to that in source 183 \citep{guo2018}. Faders can have peculiar and relatively flat extinction variations with color \citep[e.g.,][]{rodriguez2016}. However, the truly neutral color of the source 183 event makes it unique and might indicate that the occultation involves an optically thick disk.

\section{ Variable Background Stars}

The majority of the variable stars we discovered (170/289 = 59\%) are in the region of the CMD that indicates they are background to the Tr37 cluster. In general, these stars appear in deep optical images (e.g., Gaia), but have not been identified as having outstanding peculiarities. There is a slight shift to stars with similar amplitudes in $J$ and $K$ (from 54\% of the cluster members to 61\% of the background stars) and away from those with larger amplitudes at $J$ (from 26\% of the cluster members to 18\% of the background stars); only 5 (2\%) have amplitudes at $J$ more than twice those at $K$, and 3 of these are sufficiently faint that the amplitudes may have relatively large errors. This low rate of strong $J$ variability suggests a much lower fraction of strongly accreting systems in the background, as expected.  Sources 175, 204, and 302 are listed by \citet{errmann2013} as possible members of Tr 37. The first two have missing $I$ magnitudes, presumably because they are very red, which would account for why our CMD selection would have missed them. Source 302 falls on the CMD well below the Tr37 locus, calling its membership into question. 

Some of the background variables could be, for example, distant asymptotic giant branch stars. A few have interesting patterns of variability, however. We call attention to Source 69: RA = 323.521451, DEC = 57.213179 with large slow variations, Source 191: RA = 324.232413, DEC = 57.877623 which faded by a magnitude over the course of our observations,  and Source 302: RA = 324.602574 and DEC =  57.460107 that showed such chaotic variations that it must have been changing output substantially during the sequence of our individual observation sets.

\section{Conclusions}

We report the results of a two-year monitoring of $JHK$ - band variability in the Tr 37 cluster, using WFCAM on UKIRT and obtaining 21 epochs of observation. We select candidate cluster members from the $K$, $I-K$ CMD, with a check that their parallaxes from Gaia do not indicate they are foreground objects. The dividing line between background stars and cluster members was determined by the placement of reliable cluster members with accurate parallaxes from Gaia on the CMD. Only  $\lesssim 0.3$\% of the field stars in our survey are variable, compared with $65 \pm 8$\% of cluster members, indicating that our method does not mistakeningly include a significant number of field stars in our cluster sample. Because this sample is selected purely on the basis of variability and location on the CMD, it should be homogeneous and suitable for studying the general behavior of the cluster members down to very low mass objects. We use the following methods to analyze the behavior of the cluster members:

\begin{itemize}

\item{The path of the variations on the $J-H$, $H-K$ and $J-K$, $M_K$ diagrams is used to identify stars with accretion hot spots and those with variations due to extinction by dust clumps in circumstellar disks.}

\item{Hotspots show little variation in $K$ compared with larger variations in $J-K$, whereas extinction variations follow tracks of similar slope in both diagrams.}

\item{The scatter around the color-color tracks is significantly larger for stars with accretion hotspots than for those with variable extinction.}

\end{itemize}

 We reach the following conclusions about the cluster members:

\begin{itemize}

\item{Using variability to identify faint members of Tr 37, we constructed a $K$-band luminosity function for $2  < M_K < 7.5$. It drops dramatically going from the stellar range to the brown dwarf one at $M_K \sim 6.5$. Similar behavior has previously been observed for IC 348, a cluster of similar age. However,  the luminosity function for Tr37 is shifted toward brighter absolute magnitudes (by 0.3 $-$ 0.5 mag), suggesting that its K and M stars are somewhat younger than those in IC 348.}



\item{The largest variations occur in the $J$ band where a number of sources have amplitudes of $\sim$ a factor of three. Such variations are indicative of accretion events. }

\item{Eight sources have sharp, short drops in brightness suggestive of eclipses. These sources should be observed with a more rapid cadence than we used to test this possibility.}

\item{Four extremely variable sources lie at the extreme for classical T Tauri star behavior and occasionally cross into having characteristics of protoplanetary disks. They may resemble EXor variables.}

\item{Eleven sources have variability behavior consistent with varying extinction in their circumstellar disks, characterized by maximum grain sizes substantially larger than those in the general interstellar medium.}

\item{The sources with larger variability at $K$ than at $J$ and $H$ show very similar color temperatures for their variable components, close to the expected temperature for sublimation of transiently exposed dust grains. This behavior is evidence for turbulence at the inner rim of their circumstellar disks, which exposes previously shielded grains to the radiation from the star.} 

\item{One source has a long-duration dip of 0.45 mag with neutral $JHK$ color, probably due to occultation by a warp or disturbance in a possibly optically thick circumstellar disk.}

\end{itemize}

\section*{Acknowledgements}

%
We thank Watson Varricatt at UKIRT for helping us with the WFCAM observations. We thank Michael A. Read (WSA, IfA, Edinburgh),  for careful data reduction, image processing, and providing us the WFCAM data catalogs. We also thank Scott Wolk as referee for a critical reading of the paper. We used the UKIRT Wide Field Camera \citep[WFCAM;][]{casali2007} and a photometric system described in \citet{hewett2006}. The pipeline processing and science archive are described in \citet{irwin2007} and \citet{hambly2008}. 
%
We thank Roc Cutri for investigating the issues of deblending in the WISE data and alerting us to the likelihood of bad photometry for sources that are nominally reasonably well resolved but still close to each other on the sky. 
This publication makes use of data products from the Two Micron All Sky Survey, which is a joint project of the University of Massachusetts and the Infrared Processing and Analysis Center/California Institute of Technology, funded by the National Aeronautics and Space Administration and the National Science Foundation. This publication makes use of data products from the Wide-field Infrared Survey Explorer, which is a joint project of the University of California, Los Angeles, and the Jet Propulsion Laboratory/California Institute of Technology, funded by the National Aeronautics and Space Administration. When (some of) the data reported here were acquired, UKIRT was supported by NASA and operated under an agreement among the University of Hawaii, the University of Arizona, and Lockheed Martin Advanced Technology Center; operations were enabled through the cooperation of the East Asian Observatory. The Pan-STARRS1 Surveys (PS1) and the PS1 public science archive have been made possible through contributions by the Institute for Astronomy, the University of Hawaii, the Pan-STARRS Project Office, the Max-Planck Society and its participating institutes, the Max Planck Institute for Astronomy, Heidelberg and the Max Planck Institute for Extraterrestrial Physics, Garching, The Johns Hopkins University, Durham University, the University of Edinburgh, the Queen's University Belfast, the Harvard-Smithsonian Center for Astrophysics, the Las Cumbres Observatory Global Telescope Network Incorporated, the National Central University of Taiwan, the Space Telescope Science Institute, the National Aeronautics and Space Administration under Grant No. NNX08AR22G issued through the Planetary Science Division of the NASA Science Mission Directorate, the National Science Foundation Grant No. AST-1238877, the University of Maryland, Eotvos Lorand University (ELTE), the Los Alamos National Laboratory, and the Gordon and Betty Moore Foundation.

\facility{UKIRT (WFCAM)}

\begin{appendix}

In addition to the UKIRT/WFCAM data obtained for this project, we make use of the Pan-STARRS optical photometry, similar photometry by \citet{barentsen2011}, 2MASS, and WISE. This appendix discusses the methods used to adapt these data sources to our study.

\smallskip

\section{Conversion of Pan-STARRS photometry to $I_{CFH12K}$}

We have derived $I$ magnitudes by reference to the $I_{CFH12K}$ band, with an effective wavelength of 8090 \AA\ \footnote{\url{http://www.cfht.hawaii.edu/Instruments/Filters/cfh12k.html}}. We derive these magnitudes by transformation from Pan-STARRS, defined from the photometry in IC 348 by \citet{luhman2016}. 
Pan-STARRS uses $grizy$ bands for photometry in AB magnitudes, while the photometry in \citet{luhman2016} was based on different optical filters in Vega magnitudes. To reconcile the bandpass differences, we use the coordinates and $I$ band magnitudes of IC 348 members with no infrared excesses\footnote{The coefficients of magnitude conversion are slightly dependent on the input spectrum. Although no significant excess is expected at visible wavelengths, to be conservative we only use IC 348 member stars without infrared excesses to derive the transformation equation.} \citep{luhman2016} to cross-match with the Pan-STARSS DR1 astrometry and photometry  (with the subscript indicating the instrument, hereafter) The $I_{CFH12K}$ band is between those of Pan-STARRS $i_{P1}$ and $z_{P1}$ bands \citep[7520 and 8660 \AA, respectively,][]{tonry2012}. Therefore, we use the $i_{P1}$ and $z_{P1}$ photometry of IC 348 to derive the following transformation equation,
\begin{equation}\label{mag_conv}
I_{CFH12K} = 1.0291 z_{P1} + 0.2469 (i_{P1} - z_{P1}) - 0.5894
\end{equation}
The difference between Vega and AB magnitude systems is not a problem here as the transformation equation should have accounted for the zero point fluxes. 
The errors in photometric transformations are difficult to estimate; we estimate errors of 0.1 to 0.15 magnitudes for the $I_{CFH12K}$ magnitudes.
To obtain the equivalent magnitudes of the UKIRT-detected stars in the TR 37 region, we apply equation~\eqref{mag_conv} to Pan-STARRS $i_{P1}$ and $z_{P1}$ photometry for $I_{CFH12K}$ and use the transformation equation from UKIRT/WFCAM to 2MASS for the $K_{s,2MASS}$ magnitudes \citep[Equation 5d in][]{hewett2006}. We adopt an average extinction of $A_V = 1.3$ \citep{errmann2013}, corresponding to $A_{K_{s,2MASS}} = 0.148$ and $E(I_{CFH12K} - K_{s,2MASS}) = 0.491$ assuming the extinction law with $R_V = 3.1$, to correct the interstellar extinction of all stars detected by UKIRT.

\section{Identification of variations between 2MASS and 2MASS 6X Measurements}

We computed the significance (in standard deviations) of the changes between the 2MASS and 2MASS 6X measurments in each band, $J$, $H$, and $K$, and using the tabulated error estimates. A source was identified as being variable if it varied by at least 2$\sigma$ and in the same direction in at least two bands. This approach is roughly equivalent to requiring a Stetson Index $\gtrapprox$ 1, although with only two observations this parameter is probably not fully appropriate.

\section{Identification of Variability in WISE Data}

An issue with WISE is source confusion and unreliable deblending of close sources (Roc Cutri, private communication). We inspected the image of each source and flag as ``conf.'' those that appeared blended. We also found that a moderately brighter source within 15$''$ could affect the photometry, as could a source of comparable brightness within 10$''$. Both cases are also marked ``conf.'', as are any cases with sources within one beam width (6\farcs5) and bright enough to affect the results. All such sources were rejected from the further analysis. 

Nominally, with a total of $\sim$ 50 measurements in each of the two relevant WISE bands, having a $2 \sigma$ fluctuation in the same direction in both bands would already indicate a real variation with reasonable confidence. However, this is not a very conservative threshold, particularly if there is a possibility, e.g., that non-statistical fluctuations are correlated between the bands (such behavior could also ``fool'' the Stetson Index).  Given the large number of measurements, we could base the identification of variables on the scatter within the data. We computed an overall mean and standard deviation for all the measurements of a source in a given band. We then set a flag if in the same measurement set (i.e., at the same time) there was a deviation from this mean by $\ge$ 2 $\sigma$ in both bands. However, given the number of such cases, we need to calibrate the significance level associated with the nominal standard deviation. We found that the number of these flags for a given source dropped quickly from many cases with a single flag to a much lower incidence of five flags, and then persisted at similar levels to much larger numbers of flags, indicating that six and more flags were finding true variations, whereas fewer flags were contaminated by noise.  We therefore identified a source as variable if it had 6 or more of these flags, i.e., it exceeded 2 $\sigma$ from the mean in both bands in at least six measurement pairs. We identified the variations as fast if they were apparent in one or two of the 2-3 day series and slow if they appeared as a shift between the series.

\end{appendix}

\eject

\clearpage


\clearpage

\clearpage


\clearpage

\clearpage


\setcounter{table}{2}

\LongTables

\hspace{-1cm}
\begin{deluxetable}{lllccccccccccc}
\tabletypesize{\footnotesize}
\renewcommand\thetable{6}
\tablecaption{Member Candidates of Tr 37 from the CMD and Variability}
\tablewidth{0pt}
\tablehead{
\colhead {ID} &
\colhead {RA} &
\colhead {DEC} &
\colhead {$I$} &
\colhead {$ J $}  &
\colhead {$H$}  &
\colhead {$K$} &
\colhead {$\Delta J$}  &
\colhead {$\Delta H$}  &
\colhead {$\Delta K$} &
\colhead {$<S>$}    &
\colhead{var}  &
\colhead{Previous}  &
\colhead {FX (10$^{-6}$} \\
\colhead {} &
\colhead {} &
\colhead {} &
\colhead {} &
\colhead {}  &
\colhead {}  &
\colhead {} &
\colhead {}  &
\colhead {}  &
\colhead {} &
\colhead {}    &
\colhead{type\tablenotemark{a}}  &
\colhead {ID\tablenotemark{b}}  &
\colhead {ph cm$^{-2}$ s$^{-1})$}
}
\startdata

11 & 323.141836	&	57.518568	&	16.84	&	15.01	&	14.24	&	13.92	&	0.10	&	0.09	&	0.07	&	0.91	&	c	&		&				\\
13 &	323.144315	&	57.861423	&	16.17	&	13.86	&	12.93	&	12.52	&	0.11	&	0.13	&	0.18	&	0.88	&	a	&		&				\\
15 &	323.157401	&	57.650855	&	14.20	&	12.79	&	12.41	&	12.19	&	0.09	&	0.07	&	0.07	&	1.00	&	b	&		&				\\
16 &	323.159695	&	57.229836	&	16.08	&	13.81	&	12.89	&	12.48	&	0.14	&	0.10	&	0.09	&	1.22	&	c	&		&				\\
24 &	323.201657	&	57.655204	&	18.51	&	16.28	&	15.39	&	14.99	&	0.14	&	0.11	&	0.13	&	1.01	&	b	&		&				\\
25 &	323.207873	&	57.855622	&	18.04	&	15.15	&	14.09	&	13.57	&	0.13	&	0.09	&	0.11	&	1.18	&	b	&		&				\\
28 &	323.213483	&	57.642191	&	15.76	&	13.82	&	13.08	&	12.77	&	0.09	&	0.09	&	0.09	&	0.82	&	b	&		&				\\
30 &	323.240324	&	57.675013	&	17.78	&	15.63	&	14.77	&	14.39	&	0.11	&	0.11	&	0.10	&	1.18	&	b	&		&				\\
\enddata
\tablenotetext{a}{~a = variability is $>$ 30\% larger in magnitudes at $K$ than at $J$; b = variability 
is within 30\% of the same in magnitudes at $J$ and $K$; c = variability is $>$ 30\% larger at $J$
than at $K$. If the source is illustrated in the text as a variability type, the information is also provided. The relative sizes of the variations in $J$ and $K$ provide hints of the type of variations: category ``a'' is likely to vary through instabilities at the inner edge of the circumstellar disk resulting in different exposure of the disk dust; category ``c'' can arise either through extinction or accretion hot spots; in general the ratio $\Delta J$/$\Delta K$ is $<$  2 for extinction variations and $>$ 2 for hot spot ones. }
\tablenotetext{b} {Previous designations are given as \citet{errmann2013}/Sicilia-Aguilar. The latter cases are with the designation from \citet{sicilia-aguilar2006a} or by SA13 if the identification is in \citet{sicilia-aguilar2013}.}
\tablenotetext{c}{nominal error between 0.01 and 0.02 mag}
\tablenotetext{d}{also SA13}
\tablenotetext{e} {illustrated in text}

\end{deluxetable}

\clearpage



\clearpage



\begin{deluxetable}{lllccccccccc}
\tabletypesize{\footnotesize}
\renewcommand\thetable{7}d
\tablecaption{Additional Variable Sources in the UKIRT Field}
\tablewidth{0pt}
\tablehead{
\colhead {ID} &
\colhead {$RA$} &
\colhead {$DEC$} &
\colhead {$ I $}  &
\colhead {$J$}  &
\colhead {$H$} &
\colhead {$K$}  &
\colhead {$\Delta J$}  &
\colhead {$\Delta H$} &
\colhead {$\Delta K$}    &
\colhead {$<S>$}  &
\colhead {var type\tablenotemark{a}}
}
\startdata

1	&	323.087714	&	57.804996	&	16.79	&	15.26	&	14.84	&	14.62	&	0.46	&	0.42	&	0.43	&	3.81	&	b	\\
2	&	323.092506	&	57.882082	&	17.48	&	15.71	&	15.16	&	14.88	&	0.28	&	0.24	&	0.25	&	2.29	&	b	\\
3	&	323.111617	&	57.645629	&	15.31	&	13.99	&	13.58	&	13.38	&	0.08	&	0.07	&	0.07	&	0.95	&	b	\\
4	&	323.112026	&	57.780937	&	16.64	&	15.20	&	14.80	&	14.57	&	0.21	&	0.26	&	0.34	&	1.77	&	a	\\
5	&	323.121516	&	57.747625	&	19.15	&	17.40	&	16.72	&	16.44	&	0.30	&	0.14	&	0.18	&	1.15	&	c	\\
6	&	323.125681	&	57.316271	&	19.33	&	17.18	&	16.27	&	15.91	&	0.26	&	0.28	&	0.13	&	1.60	&	c	\\
7	&	323.126663	&	57.684533	&	19.44	&	17.46	&	16.79	&	16.46	&	0.30	&	0.21	&	0.25	&	1.32	&	b	\\
8	&	323.134168	&	57.650785	&	16.64	&	15.09	&	14.60	&	14.37	&	0.11	&	0.12	&	0.12	&	1.11	&	b	\\
\enddata

\tablenotetext{a}{~a = variability is $>$ 30\% larger in magnitudes at $K$ than at $J$; b = variability 
is within 30\% of the same in magnitudes at $J$ and $K$; c = variability is $>$ 30\% larger at $J$
than at $K$. The relative sizes of the variations in $J$ and $K$ provide hints of the type of variations: category ``a'' is likely to vary through instabilities at the inner edge of the circumstellar disk resulting in different exposure of the disk dust; category ``c'' can arise either through extinction or accretion hot spots; in general the ratio $\Delta J$/$\Delta K$ is $<$  2 for extinction variations and $>$ 2 for hot spot ones. }

\tablenotetext{b} {~These sources are sufficiently faint that the indication of variability might be influenced by fainter nearby sources below the detection limit, so the results for them are less reliable than for the brighter sources. In addition, the errors on their measurements are larger than the nominal 2\% for bright sources.}

\tablenotetext{c} {Listed as a possible cluster member by \citet{errmann2013}. For Sources 175 and 279, the lack of $I$ magnitudes results in their not being selected in our CMD approach, so we do not have an independent measure of whether they are cluster members. Source 335 falls far from the zone for Tr 37 members on the $K$, $I-K$ CMD, potentially calling its membership into question.} 

\end{deluxetable}

\clearpage


\clearpage

\setcounter{table}{4}

\begin{deluxetable}{llllcccccc}
\tabletypesize{\footnotesize}
\tablecaption{Summary of Variability Observed in Bright Members of Tr 37}
\tablewidth{0pt}
\tablehead{
\colhead {ID} &
\colhead {Sp.} &
\colhead {RA} &
\colhead {DEC} &
\colhead {$RI$}  &
\colhead {$JHK$}  &
\colhead {$W1W2$} &
\colhead {$W1 - W2$\tablenotemark{f}}    &
\colhead {accre-}  &
\colhead {member-}\\
\colhead {} &
\colhead {Type} &
\colhead {} &
\colhead {} &
\colhead {var.}  &
\colhead {var.}  &
\colhead {var.} &
\colhead {}    &
\colhead {tion\tablenotemark{b}}  &
\colhead {ship\tablenotemark{c}}
}
\startdata

  71-1309   &		&	21 34 09.74  &	+57 29 55.0 &		&   N		&   W1 $>$13  & --  &    &PN  \\
  74-48	&	--	&	21 34 47.30  &         +57 31 14.8	&		&  N/A	&	N& 0.34 &     & P \\
   81-541	&	K5.5	&	21 35 17.45  &	+57 48 22.3	&  	&	V	&	N& 0.49  & $0.19 $      &P \\
  64-156	&	--	&	21 35 18.04  &	+57 09 44.1 	&	&	N/A	&	conf.	& -0.06  &     &P \\
  73-472	&	K5	&	21 35 18.61  &	+57 34 09.2	&		&	N/A, V(2M)	&	S	& 0.29 &      &Y \\
  73-311	&	M1.5	&	21 35 24.51  &	+57 33 01.1	&    		&	V	&	S	& 0.51 &      &P \\
  73-71	&	K6	&	21 35 30.21  &	+57 31 16.4	&		&	N/A	&	conf.	& 0.36 &      &Y \\
  72-875	&	M0.5	&	21 35 49.75  &	+57 24 04.1	&  BP, SB	&	N/A	&    S& 0.51  & 0.11      &PN \\
\enddata
\tablenotetext{a}{Only listed in \citet{sicilia-aguilar2004}.}
\tablenotetext{b} {From \citet{sicilia-aguilar2010}.}
\tablenotetext{c}{From \citet{sicilia-aguilar2006b}: Y \& Y indicated as Y; Y \& P indicated as Y; P \& P indicated as P; Y indicated as P; P indicated as P; Y \& PN indicated as P:;Y \& N indicated as PN;  PN indicated as PN.}
\tablenotetext{d} {Indicated error 0.04 to 0.06}
\tablenotetext{e} {Indicated error 0.06 to 0.08}
\tablenotetext{f} {From ALLWISE, errors of 0.03 to 0.04 unless otherwise indicated}
\tablenotetext{g} {Very bright star (5th mag) within 90``; detection possibly contaminated by scattered light halo}
\tablenotetext{h} {Two very bright stars (6th mag) within 2`; detection possibly contaminated by diffraction spike}
\tablenotetext{i} {From \citet{morales-calderon2009}.}
\tablenotetext{j} {Very close fainter object that is integrated into the WISE PSF with little effect, but undermines the PSF fitting with the UKIRT data given its $\sim$ six times higher resolution.}
\tablenotetext{k} {A nearby source interferes with the UKIRT PSF fitting but is outside the region where the WISE deblending is not valid.}
\tablenotetext{l} {Gaia parallax suggests this object is foreground to the cluster.}

\end{deluxetable}

\clearpage

\clearpage


\setcounter{table}{5}

\begin{deluxetable}{lccccccccccc}
\tabletypesize{\footnotesize}
\renewcommand\thetable{5}
\tablecaption{Summary of Variability Observed in Members of Tr 37 from \citet{sicilia-aguilar2013}}
\tablewidth{0pt}
\tablehead{
\colhead {Source} &
\colhead {$RA_{opt}$} &
\colhead {$DEC_{opt}$} &
\colhead {$RA_{IR}$} &
\colhead {$DEC_{IR}$}  &
\colhead {$J$}  &
\colhead {$H$} &
\colhead {$K$}  &
\colhead {$Jamp$}  &
\colhead {$Hamp$} &
\colhead {$Kamp$}    &
\colhead {var.} 
}
\startdata

S13-006	&	324.093260	&	57.528118	&	324.093262	&	57.528101	&	14.644	&	13.867	&	13.409	&	0.063	&	0.065        &	         0.049	&	N	\\	
S13-013	&	324.172322	&	57.367809	&	324.172327	&	57.367830	&	15.200	&	14.511	&	14.165	&	0.138	&	0.148	&	0.140	&	V	\\	
S13-016	&	324.221700	&	57.347809	&	324.221745	&	57.347835	&	15.419	&	14.719	&	14.416	&	0.040	&	0.036	&	0.039	&	N	\\	
S13-020	&	324.246250	&	57.651722	&	324.246244	&	57.651560	&	13.459	&	12.649	&	12.180	&	0.163	&	0.240	&	0.247	&	V	\\	
S13-021	&	324.247917	&	57.526389	&	324.247785	&	57.526337	&	14.500	&	13.434	&	12.625	&	0.213	&	0.244	&	0.333	&	V	\\	
S13-022	&	324.255320	&	57.571840	&	324.255526	&	57.571777	&	14.657	&	13.863	&	13.486	&	0.113	&	0.111	&	0.121	&	V	\\	
S13-025	&	324.263400	&	57.455160	&	324.263386	&	57.455113	&	14.462	&	13.683	&	13.395	&	0.097	&	0.091	&	0.108	&	V	\\	
S13-027	&	324.282125	&	57.536389	&	324.282135	&	57.536412	&	14.622	&	13.853	&	13.459	&	0.056	&	0.066	&	0.085	&	N	\\
\enddata

\tablenotetext{a} {Probable foreground star from Gaia parallax}

\end{deluxetable}

\end{document}